\documentclass[letterpaper,twocolumn,10pt]{article}
\usepackage{usenix-2020-09}

\usepackage[available,functional]{usenixbadges}
\usepackage{tikz}
\usepackage{amsmath}
\usepackage[ruled,vlined]{algorithm2e}
\usepackage{caption}
\usepackage{graphicx}
\usepackage{subcaption}
\usepackage{circledsteps}
\usepackage{booktabs}
\usepackage{hyperref}%
\usepackage{pifont}
\usepackage[english]{babel}
\usepackage{makecell}
\usepackage{color}
\usepackage{amssymb}
\usepackage{float}
\usepackage{algorithm2e}
\usepackage{xurl}

\makeatletter
\renewcommand{\sectionautorefname}{\S\@gobble}
\renewcommand{\subsectionautorefname}{\S\@gobble}
\renewcommand{\subsubsectionautorefname}{\S\@gobble}
\makeatother

\hypersetup{
  colorlinks=true,
  linkcolor=blue!90!red,
  citecolor={red!60!black},
  urlcolor={blue!60!black}
}

\usepackage[labelformat=simple]{subcaption}

\usepackage{booktabs}
\usepackage{enumitem}

\usepackage{multirow}

\setlist{topsep=0pt, itemsep=0pt, parsep=0pt, leftmargin=1em}

\usepackage[compact]{titlesec}

\newcommand{\sys}{FalconFS}
\newcommand{\company}{Huawei}

\definecolor{darkgreen}{rgb}{0.0, 0.8, 0.0}
\definecolor{darkred}{rgb}{0.8, 0.0, 0.0}

\newsavebox{\twosubbox}

\begin{document}

\date{}

\title{\Large \bf {\sys}: Distributed File System for Large-Scale Deep Learning Pipeline}

\author{\rm Jingwei Xu\textsuperscript{1,2},\; Junbin Kang\textsuperscript{2},\; Mingkai Dong\textsuperscript{1},\; Mingyu Liu\textsuperscript{2},\; Lu Zhang\textsuperscript{2},\; Shaohong Guo\textsuperscript{2},\; \\
\rm Ziyan Qiu\textsuperscript{2},\; Mingzhen You\textsuperscript{1},\; Ziyi Tian\textsuperscript{1},\; Anqi Yu\textsuperscript{2},\; Tianhong Ding\textsuperscript{2},\; Xinwei Hu\textsuperscript{2},\; and Haibo Chen\textsuperscript{1,2}\\
  \textsuperscript{1}Institute of Parallel and Distributed Systems, Shanghai Jiao Tong University \\
  \textsuperscript{2}Huawei Technologies
} %

\maketitle

\begin{abstract}
Client-side metadata caching has long been considered an effective method for accelerating metadata operations in distributed file systems (DFSs).
However, we have found that client-side state (e.g., caching) is not only ineffective but also consumes valuable memory resources in the deep learning pipelines.
We thus propose {\sys}, a DFS optimized for deep learning pipelines with the stateless-client architecture.
Specifically, instead of performing client-side path resolution and caching, {\sys} efficiently resolves paths on the server side using \emph{hybrid metadata indexing} and \emph{lazy namespace replication}.
{\sys} also boosts server concurrency with \emph{concurrent request merging} and provides easy deployment with \emph{VFS shortcut}.
Evaluations against CephFS and Lustre show that {\sys} achieves up to 5.72$\times$ throughput for small file read/write and up to 12.81$\times$ throughput for deep learning model training.
{\sys} has been running in {\company} autonomous driving system's production environment with 10,000 NPUs for one year and has been open-sourced.

\end{abstract}
\section{Introduction}%
\label{sec:intro}
Distributed file systems (DFSs) are essential components of modern data centers.
By providing POSIX-compliant file interfaces within a unified, hierarchical directory structure, DFSs enable general access to underlying storage resources, thereby simplifying storage management and facilitating data sharing among diverse applications~\cite{3FS,GeminiFS}.
As a result, DFSs form the foundational storage layer for critical data center services, such as block storage and object storage~\cite{Lv2022InfiniFS,285788}, as well as for a broad array of applications, including data processing and analysis~\cite{Hive,Shvachko2010The-Hadoop}, high-performance computing~\cite{Li2017LocoFS,DeltaFS}, and artificial intelligence pipelines~\cite{CFS,Zhao2022Understanding,Pan2021Facebooktextquoterights}.

\begin{figure}[t]
    \centering
    \includegraphics[width=0.95\linewidth]{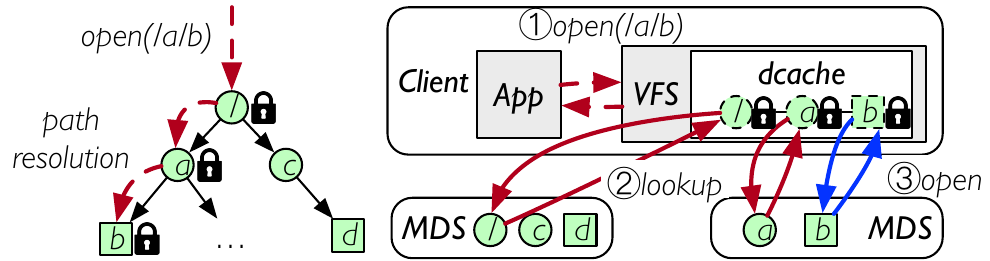}
    \caption{\textbf{Path resolution in a typical distributed file system operation.} The client checks path existence and permission by looking up each path component, which may involve multiple remote requests.
    }
    \label{fig:client-stateless}
    \vspace{-10px}
\end{figure}

However, the POSIX interface and tree-structured directory organization of DFSs are a double-edged sword, particularly in the context of deep learning (DL) workloads.
On one hand, the POSIX interface is valued for its generality and convenience, facilitating integration with existing applications and frameworks.
On the other hand, the POSIX interface and hierarchical directory organization are not well-suited for distributed environments, leading to inefficiencies in many key use cases.
Specifically, the tree-structured organization requires frequent path resolution in DFSs.
Before any file operation, the DFS client must resolve the complete path to locate the target file's inode, which involves verifying the existence and permissions of every directory along the path (\autoref{fig:client-stateless}).
As modern DFSs typically distribute directory metadata across multiple metadata servers for scalability, path resolution entails multiple round-trips between the client and metadata servers.
This leads to significant request amplification: a single file operation entails multiple network requests.

\begin{figure}[t]
    \centering
    \includegraphics[width=\linewidth]{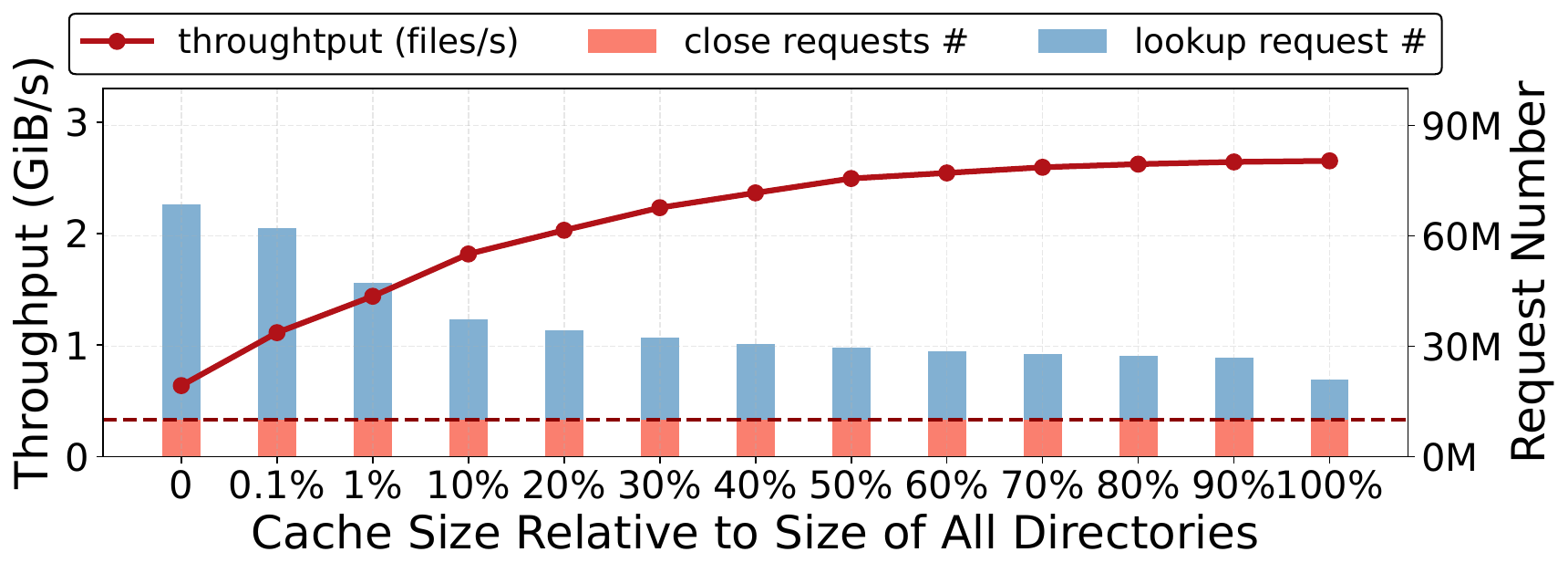}
    \vspace{-20px}
    \caption{\textbf{CephFS performance of randomly traversing 64\,KiB files in a large directory tree under different client metadata cache sizes.}
      The left y-axis represents the read throughput, and the right y-axis represents the number of requests sent to the MDSs (i.e., lookup and close). The dashed line denotes the file number. %
        }
    \label{fig:client-cache-size}
    \vspace{-15px}
\end{figure}

To mitigate path resolution overhead, existing DFSs employ client-side metadata caching, where clients maintain a local cache of directory/file metadata to avoid frequent remote lookups~\cite{nfsv4,Weil2006Ceph,lustre,2019Gluster,CFS,Ren2014IndexFS,Li2017LocoFS,Lv2022InfiniFS,Pan2021Facebooktextquoterights,BeeGFS}.
We refer to clients with client-side caching as \emph{stateful clients}, as they maintain metadata state locally.
By caching resolved paths, stateful clients reduce network round-trips for previously resolved and locally cached file operations, improving performance.

However, this \emph{stateful-client} architecture is inefficient for DL training workloads.
Unlike general-purpose workloads that exhibit strong locality and thus achieve high path resolution cache hit rates, DL training workloads demonstrate multiple traversals of massive directory trees, containing billions of directories and hundreds of billions of files in production environments, all accessed in random order.

When accessing such massive directory structures, the \emph{stateful-client} architecture faces an inherent trade-off: either consume excessive client memory to cache the large directory tree or suffer a severe performance drop due to request amplification.
To present this trade-off, we evaluate how the metadata cache size affects the performance of random file traversal in a large directory tree (shown in \autoref{fig:client-cache-size} and further explained in \autoref{sec:metadata_tax}).
Compared to a cache that can hold all directories, a cache only 10\% as large results in a 32.5\% throughput reduction, primarily because the lookup requests increase by 1.50$\times$.
The dilemma is exacerbated by the fact that a cache that can hold 10\% of the directories is prohibitively expensive in production, considering the large directory tree and the number of clients.
Even when the cache can hold 90\% of the directories, an average of 1.70 network hops (i.e., the lookup requests in \autoref{fig:client-cache-size}) are required for a file \emph{open}.

In this paper, we propose the \emph{stateless-client} architecture, which achieves one-hop access for most file operations in DL workloads while requiring no client-side caching.
The core difference is that the stateless client shifts path resolution to the server side.
However, to realize the stateless-client architecture for an efficient DFS, two key problems remain to be addressed.
First, to enable one-hop access for most file operations, the client needs an approach to \emph{finding the correct metadata server that stores the target file's inode}.
Second, as the path resolution is now on the server side, \emph{each metadata server should be able to resolve paths locally}, without involving additional network hops when processing clients' file operation requests.

We realize the \emph{stateless-client} architecture in {\sys}, a POSIX-like high-performance DFS designed for DL workloads.
{\sys} proposes \emph{hybrid metadata indexing} to address the first key problem of locating the appropriate metadata server (\autoref{sec:filename-direct-indexing}).
This approach leverages filename hashing to place file inodes to metadata servers while achieving load balance with selective redirections.

To address the second problem of resolving path locally, {\sys} proposes \emph{lazy namespace replication}, replicating the whole namespace (i.e., the directory tree) to all metadata servers (\autoref{sec:lazy-replication}).
To amortize synchronization overhead, namespace updates are lazily synchronized, and an invalidation-based mechanism is adopted for concurrency control.

{\sys} further adopts \emph{concurrent request merging} to amortize operation overhead and improve concurrency (\autoref{sec:request-merging}) and \emph{VFS shortcut} to keep the stateless-client architecture compatible with Linux VFS for easy deployment (\autoref{sec:lookup-pass-through}).

Evaluations show that {\sys} completely eliminates request amplification and improves performance of large directory tree random traversal by up to 4.72$\times$ over CephFS~\cite{Weil2006Ceph} and up to 3.34$\times$ over Lustre~\cite{lustre}.
End-to-end experiments show that {\sys} improves the throughput of DL model training by up to 11.81$\times$ and 1.23$\times$ over CephFS and Lustre, respectively.
{\sys} has been deployed in {\company}'s AI clusters with 10,000 NPUs for data labeling and model training of the autonomous driving solution for one year.

Our main contributions are as follows.
\begin{itemize}
    \item Analyzing IO characteristics of deep learning workloads, pointing out the inefficiency of the traditional stateful-client architecture.
    \item Proposing the stateless-client architecture and addressing the challenges of designing an efficient DFS.
    \item Building {\sys}, a high-performance DFS designed for DL workloads, and a thorough evaluation on it.
\end{itemize} 

{\sys} is open-sourced at  \url{https://github.com/falcon-infra/falconfs} and \url{https://gitee.com/openeuler/FalconFS}.

\section{DL Pipelines: IO Patterns and Challenges}
\label{sec:background}

\subsection{Deep Learning Pipeline}
\label{sec:pipeline}

\begin{figure}[t]
    \centering
    \includegraphics[width=\linewidth]{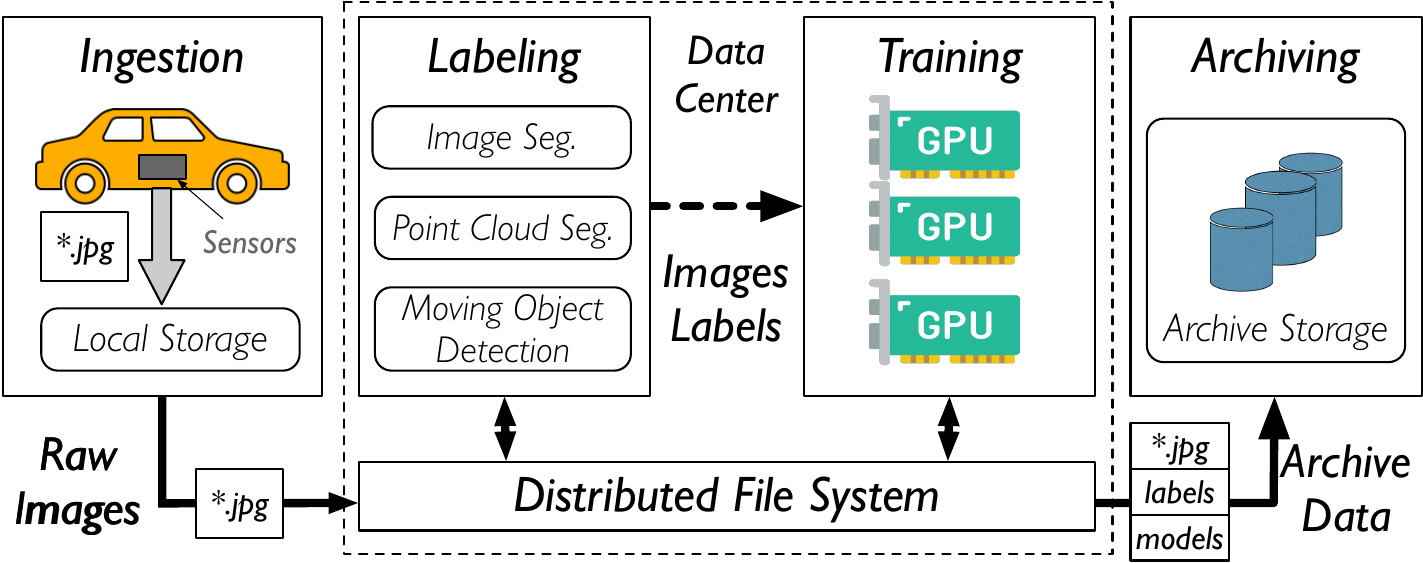}
    \vspace{-20px}
    \caption{\textbf{Overview of a DL Pipeline for Autonomous Driving.}}
    \label{fig:pipeline_architecture}
    \vspace{-15pt}
\end{figure}

Deep learning (DL) models are widely employed in autonomous driving, computer vision, and big data analysis.
To meet evolving quality demands, training pipelines have been developed to continuously retrain models using large, iteratively updated datasets.

\autoref{fig:pipeline_architecture} illustrates the architecture of the deep learning (DL) training pipeline employed in {\company}'s autonomous driving (AD) systems. The pipeline consists of four key stages:
The \emph{ingestion} stage collects raw data from real-world environments.
The \emph{labeling} stage generates labels for the raw data with a sequence of model inference tasks, including moving objects detection, lane detection, traffic sign detection, etc.
Then, the \emph{training} stage uses a labeled dataset to train the target model for vehicle deployment.
In the final \emph{archiving} stage, the dataset is moved into low-cost storage systems such as cloud data lakes for future reference~\cite{Zaharia2021LakehouseAN,tf.data,Zhao2022Understanding,Cachew}.
Similar architectures also exist in other DL workloads~\cite{Zhao2022Understanding,netapp-dp,netapp-ad}.

\subsection{Workload Patterns in DL pipelines}
\label{subsec:pattern}

We take the DL pipeline for autonomous driving in {\company} as an example to present its unique workload patterns.

\paragraph{Numerous small objects in large directories.}
The autonomous driving pipeline consumes multimodal data, including images, point clouds, etc.
During labeling and training, these objects are stored as individual files, whose size ranges from a few KiB to a few MiB, mostly within 256KiB.
In production, an in-flight dataset scales up to hundreds of petabytes and is composed of over 300 billion small files.
These files are grouped into directories by timestamps, vehicle ID, camera ID, etc., forming a directory tree with billions of directories and large directory sizes.
The huge number of files and directories stresses the DFS's metadata scalability and performance.

\paragraph{Random file traversal.}
In the training stage, tasks access the dataset in a \emph{traversal} and \emph{random} manner.
In particular, each file is accessed exactly once in each training epoch, and the access order is random.
Such a random access pattern is unfriendly to client caching, which we further discuss in \autoref{sec:metadata_tax}.

\paragraph{Burst file access.}
In the labeling stage, the inference tasks read and write files from/to {\sys} in a pipeline, involving massive small file IO and directory lists.
To fully utilize the GPU's parallelism, data objects are accessed and processed in batches, resulting in burst file access in the same directory, which can lead to instantaneous load imbalance on the metadata servers and downgrade the performance, which we further discuss in \autoref{sec:server-concurrency}.

\paragraph{Tight resource budget.}
CPU and memory are scarce resources for computing nodes.
Training tasks perform data augmentation on CPUs, consuming significant CPU cycles and memory resources.
In production, CPU for data augmentation is often the bottleneck, and it is typical to store and reuse intermediate results in memory to reduce CPU load~\cite{Cachew,Revamper}.
Therefore, the resources available to DFS clients are limited.

\subsection{Challenge 1: Lookup Tax}
\label{sec:metadata_tax}

For small-file intensive workloads like DL pipelines, metadata performance often becomes the bottleneck.
To accelerate metadata operations, most DFSs employ a \emph{stateful client} architechture that caches metadata on the client side~\cite{nfsv4,Weil2006Ceph,lustre,2019Gluster,CFS,Ren2014IndexFS,Li2017LocoFS,Lv2022InfiniFS,Pan2021Facebooktextquoterights,BeeGFS}.
However, this approach is inefficient for DL training workloads, as their large directory working sets are inefficient to cache, leading to the following dilemma.

\paragraph{Challenge 1:}
Large working sets in deep learning training tasks create a dilemma between performance degradation due to \emph{request amplification} and excessive \emph{client memory consumption} required for caching directory metadata.

To illustrate this dilemma, we replay a trace of Resnet-50 model training on a small dataset, comprising 10 million 64KiB files in 1 million 7-level directories, stored in a CephFS instance with four MDSs and twelve OSDs\footnote{Ceph MDS (metadata server daemon) and OSD (object storage daemon) can be considered as CephFS's metadata server and data node, respectively.}.
A total of 512 IO threads iterate through the files in random order.
As shown in \autoref{fig:client-cache-size}, the client cache size significantly impacts both read throughput and the number of remote requests generated.

The experiment shows that the read throughput is highly sensitive to the client's metadata cache size.
Compared to a cache that can hold 10\% of the directories, a cache that can hold all directories achieves 1.46$\times$ higher throughput.
Furthermore, throughput increases significantly as the cache size grows from 10\% to 100\%, indicating that optimal performance requires allocating sufficient memory to cache all directories. Otherwise, performance degrades proportionally with reductions in cache size.

The DL training workload's sensitivity to cache size comes from its random access nature.
During each epoch, training tasks access every data object exactly once in a random order. 
When all directories are cached, all directory lookups are served by the cache, reducing each file \emph{open} operation to a single metadata lookup request. However, with a small cache size, the LRU policy preferentially retains near-root directories, while the hit rate of last-level directories --- which constitute 90\% of accesses in the experiment —- is proportional to the cache size. 
As shown in \autoref{fig:client-cache-size}, smaller caches correlate with increased lookup requests due to cache misses. This effect causes request amplification: each file \emph{open} operation triggers multiple lookup requests to the metadata server, bottlenecking read throughput.

In production environments, caching a significant portion of working-set directories proves prohibitively expensive. For instance, a production cluster in {\company} can scale over 1000 client nodes, and a typical production dataset contains billions of directories.
Given that in Linux VFS, caching a directory takes 800 bytes (608 bytes for inode and 192 bytes for dentry), caching 10\% of 1 billion directories on each node would require 80\,GiB per node and 80\,TiB in total, which is prohibitively expensive.

\subsection{Challenge 2: Transient Skewness}
\label{sec:server-concurrency}
\begin{figure}[t]
    \centering
    \begin{minipage}[t]{.495\linewidth}
        \includegraphics[width=\linewidth]{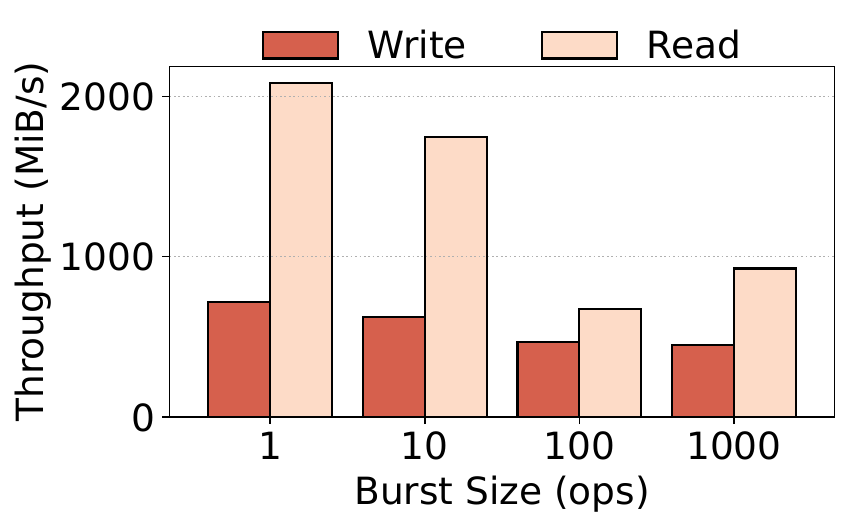}
        \vspace{-20px}
        \subcaption{\textbf{Throughput.}}
        \label{fig:directory-burst-a}
    \end{minipage}
    \begin{minipage}[t]{.495\linewidth}
        \includegraphics[width=\linewidth]{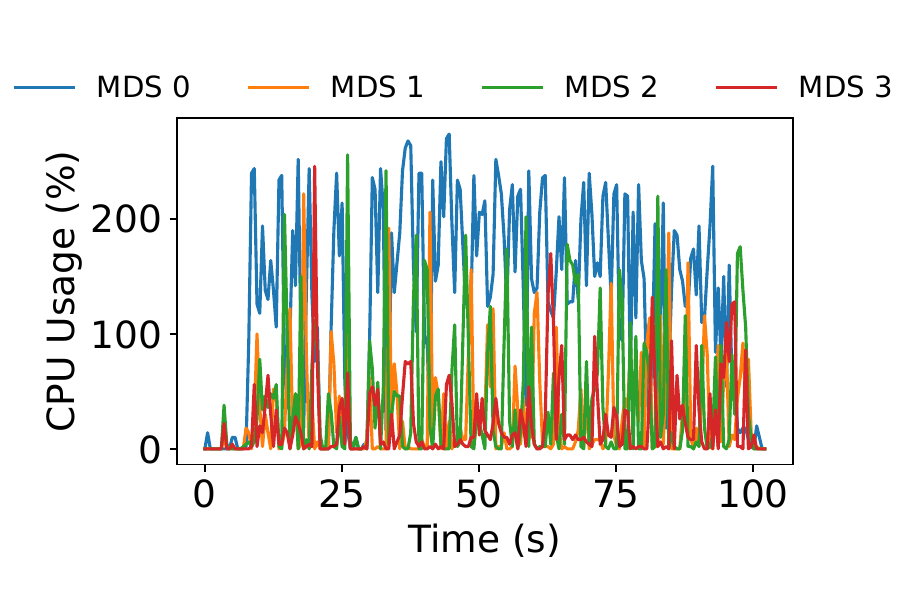}
        \vspace{-20px}
        \subcaption{\textbf{Load Variance.}}
        \label{fig:directory-burst-b}
    \end{minipage}
    \vspace{-12px}
    \caption{\textbf{CephFS performance of accessing 64KiB files with different burst size.}
        Large burst size leads to load imbalance and performance degradation.
        \autoref{fig:directory-burst-b} shows the load variance of the MDSs when performing read operations with burst size 100.
        }
    \label{fig:directory-burst}
    \vspace{-15px}
\end{figure}

During the labeling stage, inference tasks scan and load/store data objects in a per-directory manner, accessing all the files in one directory and then another.
This per-directory IO pattern causes transient load imbalances, which limit performance scalability.
We illustrate this issue by performing per-directory file access on a CephFS cluster with four MDSs and twelve OSDs and present the results in \autoref{fig:directory-burst-a}.

We observe that when the directory size exceeds the IO parallelism of the tasks, the DFS's performance degrades.
Analysis of load distribution (\autoref{fig:directory-burst-b}) shows that at a directory size of 100, the MDSs experience severe load imbalance during read operations. This issue arises because CephFS tends to store metadata for files within the same directory together; consequently, burst operations on files in the same directory can congest a single MDS, leading to performance bottlenecks. 
Similar patterns also exist in other DFSs like \cite{BeeGFS,Ren2014IndexFS,Lv2022InfiniFS,CFS,Pan2021Facebooktextquoterights}.

\paragraph{Challenge 2:} 
Concurrent operations on files within the same directory suffer from MDS congestion, hindering performance scalability.

\section{Proposal: DFS with Stateless Client}

Given the challenges brought by DL workloads, we propose the \emph{stateless-client} architecture for DFS under DL workloads.
The \emph{stateless-client} architecture abandons the cache on the client side and moves path resolution to the server side.
The proposal is based on the following three observations.
(1) A typical DL cluster usually has far more clients than metadata servers. In {\company}'s deployment environment, the ratio of clients to servers exceeds 40:1.
Considering that all clients share the same dataset in DL pipelines, a server-side dentry can serve more clients than a client-side cache.
(2) Stateful-client DFSs typically use VFS dcache and inode cache to cache directory attributes~\cite{Weil2006Ceph,lustre,juicefs,BeeGFS}, which takes 800 bytes for each directory. On the server side, a directory entry can be maintained in a custom format that takes less than 100 bytes.
(3) The server has more memory resources than the clients due to the tight resource budget of the computing node, as we described in \autoref{subsec:pattern}.
However, to realize the stateless-client architecture for an efficient DFS, two key problems remain. %

\paragraph{How to find the right server?}
First, the client needs an approach to \emph{finding the correct metadata server that stores the target file's inode}.
Since DFS usually partitions all inodes to multiple metadata servers for scaling out, only the server containing the file's inode can complete its file operations.
DFS thus needs to maintain the path-to-server mapping (i.e., the indexing), which is usually implemented in two ways.
\begin{itemize}
    \item \emph{Path-walk indexing.}
    Most DFSs use indexing methods that are related to the parent directories, for example, using the parent directory ID as (part of) the key~\cite{Li2017LocoFS,BeeGFS,Lv2022InfiniFS,CFS,Pan2021Facebooktextquoterights} for hashing, or indexing files in different directories with different hashing rules~\cite{2019Gluster,Ren2014IndexFS,lustre}.
    These approaches require a full-path walk to resolve inode location, and need caching directory entries on clients to achieve better performance, suffering the trade-off in \autoref{sec:metadata_tax}.
    
    \item \emph{Full-path hashing}~\cite{Giraffa,CalvinFS} determines the target inode location by hashing the full path.
    It does not require cache, but makes directory \emph{rename} prohibitively expensive due to the relocation of all inodes in the subtree.
\end{itemize}

However, none of these indexing approaches meet the requirements of an efficient DFS with the stateless-client architecture, including no client cache, no extra network hops, and practical support of directory renames.
Nevertheless, these approaches inspire us to build a hybrid indexing of hashing and path-walking to solve the problem.

\paragraph{How to resolve path locally?}
Second, as the path resolution is now on the server side, \emph{each metadata server should be able to resolve the path locally}, without involving another network hop when processing the client's file operation request.
Since DFS scatters directory entries across multiple metadata servers for scaling out, the dentry information of components in a path is also scattered across multiple servers.
A server resolving the path thus needs to communicate with other servers to complete the resolution, which can be expensive.

Considering the characteristics of DL workloads (random file traversal) and DL clusters (high client-server number ratio and rich server memory resource), we thus propose to replicate the whole directory tree (i.e., the namespace) on all metadata servers, so that each server can resolve path locally.

We build {\sys} with the \emph{stateless-client} architecture, which eliminates client-side caching while providing one-hop access for most operations in DL workloads with \emph{hybrid metadata indexing} and \emph{lazy namespace replication}.

\section{System Design}

\subsection{{\sys} Overview}

\begin{figure}[tb]
    \centering
    \includegraphics[width=0.9\linewidth]{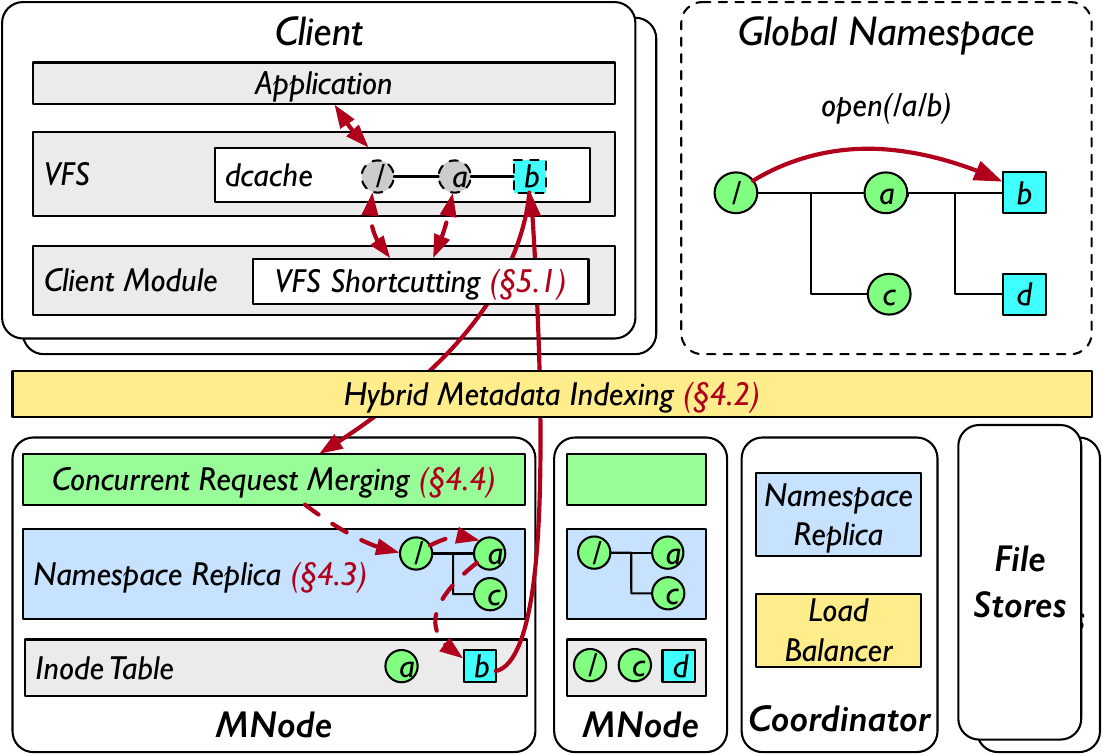}
    \vspace{-5px}
    \caption{\textbf{Architecture of {\sys}}}
    \label{fig:architecture}
    \vspace{-10px}
\end{figure}

\begin{table}[tb]
    \centering
    \caption{\textbf{Scheme of {\sys}'s metadata.}}
    \label{tab:metadata-scheme}
    \vspace{-10px}
    \begin{tabular}{llll}
    \toprule
                    & \textbf{Key}& \textbf{Value} & \textbf{Partition by}  \\
    \midrule
    dentry & pid, name & id, perm. & replicated       \\
    inode  & pid, name & id, attr & \autoref{sec:filename-direct-indexing}   \\
    \bottomrule
    \end{tabular}
    \vspace{-15px}
\end{table}

\paragraph{Architecture.}
\autoref{fig:architecture} shows the {\sys} architecture, consisting of \emph{Client Module}, \emph{MNode}, \emph{Coordinator}, and \emph{File Store}.

{\sys}'s \emph{client module} is a kernel module that provides POSIX interfaces through Linux VFS.
Unlike most Linux file systems, {\sys} client avoids performing path resolution at the client side.
Specifically, {\sys} client shortcuts VFS path walk (\autoref{sec:lookup-pass-through}) and forwards
the operation request with full path to MNodes according to hybrid metadata indexing (\autoref{sec:filename-direct-indexing}).

Metadata nodes (\emph{MNode}) are PostgreSQL databases~\cite{PostgreSQL} with customized extensions.
We manage metadata for {\sys} in the extensions using the database's table and transaction management, B-link tree index, xlog (write-ahead logging), and primary-secondary replication.
Each MNode maintains a lazily synchronized directory tree structure (namespace replica) and a partition of file attributes (inodes).
Leveraging the merits of central location as servers, MNode merges concurrent requests (\autoref{sec:request-merging}) to de-duplicate shareable execution processes for higher throughput.

The central \emph{coordinator} is responsible for managing namespace changes (\autoref{sec:lazy-replication}).
It also runs a load balancing algorithm to balance inode distribution across MNodes (\autoref{sec:filename-direct-indexing}).

\emph{File Store} is a distributed block storage system that stores file data.
File chunks are distributed across a set of file store nodes that use local file systems on SSDs to store data.

\paragraph{Replicated directory namespace}
{\sys} replicates the file system directory structure across all MNodes, enabling each MNode to resolve file paths and check permissions locally.
The namespace replica contains entries of directory attributes required for path resolution, i.e., \emph{dentries} in \autoref{tab:metadata-scheme}, and does not include file attributes.
With one billion directories, the storage footprint for the namespace replica is less than 100\,GiB per MNode, which is acceptable.

\paragraph{Sharded file metadata}
In contrast to directories, we distribute all the file metadata, i.e., \emph{inodes} in \autoref{tab:metadata-scheme}, across the metadata servers by hybrid metadata indexing for scaling up metadata capacity and throughput.

\subsection{Hybrid Metadata Indexing}
\label{sec:filename-direct-indexing}

In this section, we present hybrid metadata indexing (\autoref{subsubsection:hybrid-indexing-methods}) and how to maintain load balance (\autoref{sec:scheduling}).

\begin{figure}[t]
    \centering
    \includegraphics[width=\linewidth]{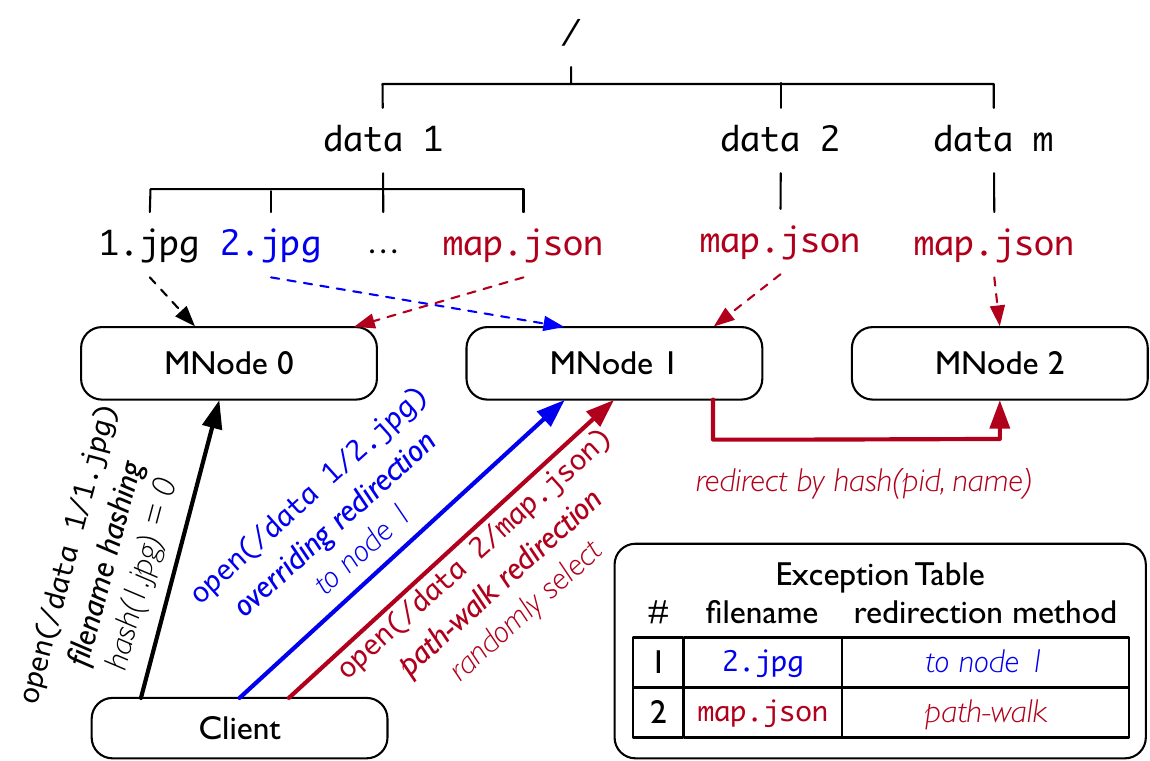}
    \vspace{-15px}
    \caption{\textbf{Hybrid Metadata Indexing.}}
    \label{fig:hybrid}
    \vspace{-20px}
\end{figure}

\subsubsection{Hybrid Indexing Methods}
\label{subsubsection:hybrid-indexing-methods}
Hybrid metadata indexing adopts filename hashing in the common case.
However, for generality, hybrid metadata indexing considers situations where filename hashing may bring uneven inode distribution, and adopts two fallback mechanisms to mitigate the problem.
\autoref{fig:hybrid} presents an overview.

\paragraph{Common case fast: filename hashing.}
To accelerate the path resolution, {\sys} adopts \emph{filename hashing} as the indexing method for most cases.
With \emph{filename hashing}, all files are placed via the hashing of the filename.
Such an indexing method is easy and efficient --- it requires no client state and supports efficient rename.
However, \emph{filename hashing} cannot guarantee that the files are distributed evenly among different servers, potentially causing load-imbalance issues.

Fortunately, we find that DL workloads' \emph{large directory size} facilitates the filename hashing, significantly reducing the occurrence of load imbalance.
Specifically, DL datasets usually have a large directory size, which can be multiple times larger than the number of MNodes.
We analyze {\company}'s in-production datasets and popular open-sourced datasets that contain more than ten thousand files~\cite{imagenet-object-localization-challenge,Geiger2012CVPR,Cordts2016Cityscapes,liu2015faceattributes,WahCUB_200_2011},
finding their directory size ranging from several hundred to hundreds of thousands.
According to the law of large numbers, given such large directory sizes, the distribution of files in each directory to MNodes is likely to be uniform. %
Consequently, the entire namespace is statistically uniformly distributed as the superposition of the distributions for each directory.
This forms the basis of using \emph{filename hashing} as the indexing method for the common cases in DL workloads.

\paragraph{Corner cases correct: selective redirection.}
However, there are situations where filename hashing can bring uneven file distribution.
(a) \emph{Hot filenames}: The naming convention of applications can cause certain filenames to be more frequent than others.
(b) \emph{Hash variance}: The number of unique filenames is not far more than that of MNodes, which can lead to unbalanced distribution due to hash variance.

To handle these corner cases, {\sys} adopts \emph{selective redirection} as a complement.
For this, 
{\sys} maintains a shared data structure --- \emph{exception table}, which specifies which and how filenames should be redirected.
There are two kinds of redirection, targeting hot filenames and hash variance, respectively.

\begin{itemize}
    \item \emph{Path-walk redirection.}
    To place files with hot filenames, {\sys} calculates the hashing value by not only the filename, but also its parent directory ID.
    Thus, even if a hot filename exists in many directories, the corresponding files are placed on different MNodes.
    When clients identify a file marked for path-walk redirection in the exception table, they send requests to random MNodes.
    The receiving node utilizes its local namespace replica to walk the path, obtains the parent directory ID, and forwards the request to the target node determined by hashing both the filename and the parent directory ID.

    \item \emph{Overriding redirection.} When hash variance causes uneven filename distribution across nodes, {\sys} can reassign selected filenames to designated nodes, shifting load from overloaded to underutilized nodes.
    These placement overrides are maintained in the exception table.
    Clients encountering files marked for such redirection in the exception table send requests directly to their designated MNodes.

\end{itemize}

{\sys} maintains copies of the exception table on each MNode, clients, and the coordinator.
The coordinator updates the exception table according to the scheduling policy detailed in \autoref{sec:scheduling}.
Once the exception table is updated, the latest table is eagerly pushed to all MNodes, and clients lazily fetch the updates from MNodes when they get responses to operation requests.
The lazy fetching mechanism creates a time window where clients may operate with stale exception tables and direct requests to incorrect nodes.
However, MNodes validate all requests by checking their local exception table and forward misdirected requests to the proper destinations.

A constant number of exception table entries is sufficient to balance the distribution of inodes for arbitrary directory structures. 
We provide a theoretical analysis in \autoref{sec:theoretical-analysis}.

\subsubsection{Statistical Load Balancing}
\label{sec:scheduling}

The coordinator uses the statistics periodically reported by MNodes to make rebalancing decisions.

\paragraph{Statistics.}
Each MNode periodically reports its local inode count and the most frequent $O(nlogn)$ local filenames with their occurrence counts, where $n$ is the number of MNodes.

\paragraph{Load balancing algorithm.}
The coordinator leverages an algorithm to maintain load balance across nodes.
The algorithm aims to keep each node's inode count below $(\frac{1}{n} + \epsilon)$ of the total inode count, while minimizing the size of the exception table.
$\epsilon$ is a parameter specified by the user used to control the effect of load balancing.

When the coordinator detects a load imbalance using statistics reported by MNodes, it conducts the load balancing algorithm as follows.

\begin{enumerate}
    \item \emph{Identify the most and least loaded nodes}.
    Let $N_{max}$ and $N_{min}$ denote the nodes with the highest and lowest inode counts, respectively, where the inode counts are denoted as $\langle{N_{max}}\rangle$ and $\langle{N_{min}}\rangle$.
    \item \emph{Select the most frequent filename in $N_{\text{max}}$} as $F$.
    \item \emph{Redistribute $F$}. 
    The coordinator selects from the two redirection methods to approach load balancing.
    If we use \emph{path-walk redirection}, assuming it evenly distributes all files with filename $F$ to all nodes, the inode numbers of $N_{max}$ and $N_{min}$ will be $ \langle{N_{max}}\rangle - \frac{n-1}{n}|F|$ and $ \langle{N_{min}}\rangle + \frac{1}{n}|F|$, respectively,  where $|F|$ represents the number of files named $F$ in node $N_{max}$.
    If using \emph{overriding redirection} for $F$, we will transfer all $|F|$ files from $N_{max}$ to $N_{min}$, yielding the inode numbers of $\langle{N_{max}}\rangle - |F|$ and $ \langle{N_{min}}\rangle + |F|$, respectively.
    The coordinator chooses the method that minimizes the maximum inode count.
    After the method is chosen, the redirection entry is inserted into the exception table, and the corresponding files are migrated among nodes.
    To ensure metadata consistency, access to the corresponding inodes is temporarily blocked during the migration.
    
    \item \emph{Repeat the procedure until no further imbalance is detected.}
\end{enumerate}

Moreover, the coordinator periodically attempts to shrink the exception table to reduce redirection overhead.
Specifically, it iterates all path-walk redirection entries in random order, and removes the entry if removing it does not lead to load imbalance.
It then checks and removes overriding redirection entries similarly.

\subsection{Lazy Namespace Replication}
\label{sec:lazy-replication}

\begin{figure}[t]
    \centering
    \begin{minipage}{0.39\linewidth}
        \centering
        \includegraphics[width=0.7\linewidth]{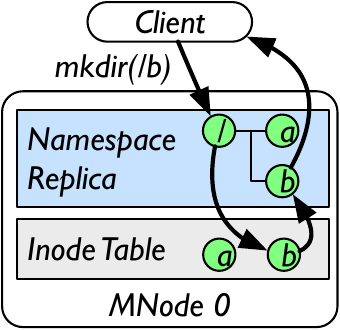}
        \subcaption{\textbf{Directory creation.}}
        \label{fig:replication-a}
    \end{minipage}
    \begin{minipage}{0.6\linewidth}
        \centering
        \includegraphics[width=0.7\linewidth]{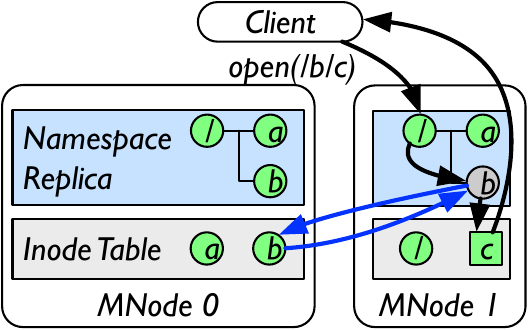}
        \subcaption{\textbf{Directory lookup and miss handling.}}
        \label{fig:replication-b}
    \end{minipage}
    \begin{minipage}{\linewidth}
        \centering
        \includegraphics[width=0.7\linewidth]{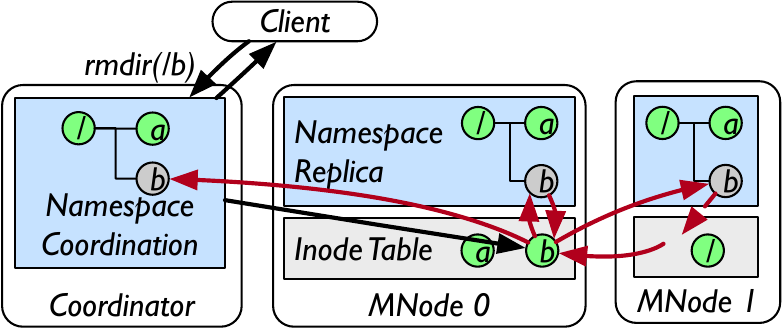}
        \subcaption{\textbf{Directory change permission, removal and rename}}
        \label{fig:replication-c}
    \end{minipage}
    \vspace{-10px}
    \caption{\textbf{Namespace Synchronization.} Blue arrows represent \emph{remote lookup} and red arrows represent \emph{invalidation}.}
    \label{fig:replication}
    \vspace{-15px}
\end{figure}

Lazy namespace replication is another key enabler of the stateless-client architecture.
{\sys} maintains a consistent but not necessarily complete namespace replica on each MNode and the coordinator, enabling local path resolution.
To reduce the overhead of maintaining consistency across all replicas, modifications to the namespace are lazily synchronized, and an invalidation-based mechanism is adopted, inspired by cache coherence protocols and Hermes~\cite{Hermes}.
The design is guided by two principles:
\begin{itemize}
    \item \emph{Delaying synchronization until access.}
    Directory creation is critical for DL dataset initialization, so the performance is important.
    Eagerly replicating directory creation across all MNodes would require expensive and unscalable two-phase commits (2PC).
    Instead, {\sys} amortizes the overhead by deferring synchronization until access.
    \item \emph{Using invalidation as lightweight locking.}
    When operations modify directory structures or permissions, concurrent operations in the sub-tree under the modified directory must be blocked for consistency.
    We invalidate the corresponding replica entry on all nodes for this instead of using traditional two-phase locking, saving a round of request broadcast.
\end{itemize}
We will then introduce how the namespace replicas are maintained during related operations, including creating/removing a directory, changing the permissions, and renaming.

\paragraph{Creating a directory.}
\autoref{fig:replication-a} shows an example of how to create the directory \emph{/b}.
The client calculates the location of \emph{/b} via the hybrid metadata indexing, and then sends a \emph{mkdir} request to the corresponding MNode, i.e., $MNode_0$.
Upon receiving the request, $MNode_0$ resolves the path by querying its local namespace replica to check the path existence and permissions.
It also checks its inode table to confirm \emph{/b} does not already exist.
Once all checks are passed, $MNode_0$ creates an inode for \emph{/b} in the inode table, adds a dentry for \emph{/b} to its local namespace replica, and responds to the client.

Note that to retain the efficient single-hop processing of \emph{mkdir}, $MNode_0$ does not proactively broadcast the new dentry to namespace replicas on other MNodes.
Instead, the dentry information is fetched by other namespace replicas on demand.
Specifically, when an MNode queries its local namespace replica for path resolution and finds a missing dentry, it fetches the information from the owner MNode calculated via the hybrid metadata indexing.

\autoref{fig:replication-b} shows an example.
When $MNode_1$ resolves the path \emph{/b/c}, it finds the dentry for \emph{/b} missing in its local namespace replica.
It then sends a \emph{lookup} request to $MNode_0$ (i.e., the owner MNode) to fetch the missing dentry to complete the path resolution and continues subsequent processing.

\paragraph{Removing a directory.}
The centralized coordinator is responsible for removing a directory and invalidating corresponding dentries from all namespace replicas.

In the example shown in \autoref{fig:replication-c}, the client sends an \emph{rmdir(/b)} request to the coordinator.
The coordinator acquires shared locks on all ancestor directories and an exclusive lock on the target directory to ensure path validity during execution.
It then forwards the request to the directory inode's owner, i.e., $MNode_1$ in the figure.
$MNode_1$ locks \emph{/b}'s inode to block subsequent \emph{lookup} requests and broadcasts an \emph{invalidation} request to other MNodes.
Upon receipt, each MNode invalidates its local dentry of the target directory if it exists in the local namespace replica.
Each MNode searches its inode table for entries whose key's \emph{pid} equals \emph{/b}'s ID (i.e., \emph{/b}'s children), and responds to $MNode_1$ the existence of any children.
$MNode_1$ aggregates these responses.
If \emph{/b} has no children, $MNode_1$ deletes the inode and notifies the coordinator, which will release locks and respond to the client.
Otherwise, $MNode_1$ returns \emph{-ENOTEMPTY} to abort the \emph{rmdir}.

\paragraph{Changing permissions.}
Operations altering file permissions are also handled by the centralized coordinator in a similar approach.
The difference is that the owner MNode will broadcast the \emph{invalidation} requests and change the permission in its inode table.

\paragraph{Rename.}
{\sys} ensures \emph{rename} consistency by employing the central coordination via conventional two-phase locking and two-phase commit protocols.
The client will send \emph{rename(A, B)} request to the coordinator, who will acquire locks and conduct checks on whether the rename can proceed.
Once all checks are passed, the coordinator broadcasts to invalidate dentries of path $A$ and transfers the inode of $A$ to the MNode who is responsible for path $B$.

\paragraph{Locking and conflict resolving.}
{\sys} leverages namespace replicas to coordinate concurrent requests.
When a server (the coordinator or an MNode) processes a request, it first resolves the path component by component in its local namespace replica.
Dentry locks are acquired during the resolution.
To improve parallelism, the coordinator acquires shared locks on intermediate dentries and an exclusive lock on the last component, while MNodes acquire shared locks on all dentries.
If a dentry is missing in the local namespace replica, the server locks the dentry after it is retrieved from its owner MNode.
Thus, concurrent requests on the same server are serialized by these locks.
Concurrent requests on two different MNodes will not incur a data race and thus can be executed in parallel.
The last case is a request being processed on the coordinator ($Req_C$) and a request on an MNode ($Req_M$).

For simplicity and without loss of generality, we assume $Req_C$ is removing \emph{/a/b}, whose inode is on $MNode_C$, and $Req_M$ is opening \emph{/a/b/c} on $MNode_M$.
During processing $Req_C$, $MNode_C$ will lock \emph{/a/b}'s inode and broadcast to invalidate the dentries of \emph{/a/b} in all namespace replicas.
To handle the invalidation request, an MNode will first lock the corresponding dentry and then mark it as \emph{invalid}.
Then there are two possible cases on $MNode_M$.

The first case is that $Req_M$ already holds the lock of \emph{/a/b} when the \emph{invalidation} arrives. In this case, the invalidation will be blocked until $Req_M$ completes, thus $Req_C$ is serialized to happen after $Req_M$.

The second case is that the invalidation is processed before $Req_M$ locks \emph{/a/b}.
In this case, $MNode_M$ will find the \emph{/a/b} dentry invalid during path resolution of $Req_M$.
$MNode_M$ then sends a \emph{lookup} request to retrieve the dentry from $MNode_C$, who owns the \emph{/a/b} inode.
On $MNode_C$, the lookup will acquire a shared lock of the \emph{/a/b} inode, which has already been locked by $Req_C$.
Thus, $Req_M$ will be blocked until $Req_C$ completes, forming a correct serialization of the two requests.
Note that when processing the invalidation request, $MNode_M$ will discard all \emph{lookup} responses whose requests are issued before the invalidation is received.

\paragraph{Discussion.}

Lazy namespace replication introduces one remote lookup for access to a non-existent path.
However, in DL workloads, such negative access is rare.

\subsection{Concurrent Request Merging}
\label{sec:request-merging}

To further optimize the performance, {\sys} leverages the advantages of stateless-client architecture and adopts concurrent request merging to scale up per-MNode throughput for higher metadata performance.
In large-scale deep learning clusters, thousands of compute nodes generate concurrent requests.
This presents an opportunity to batch request handling opportunistically, amortizing per-operation overhead --- particularly lock contention and write-ahead logging costs.

\begin{figure}[t]
    \centering
    \includegraphics[width=0.8\linewidth]{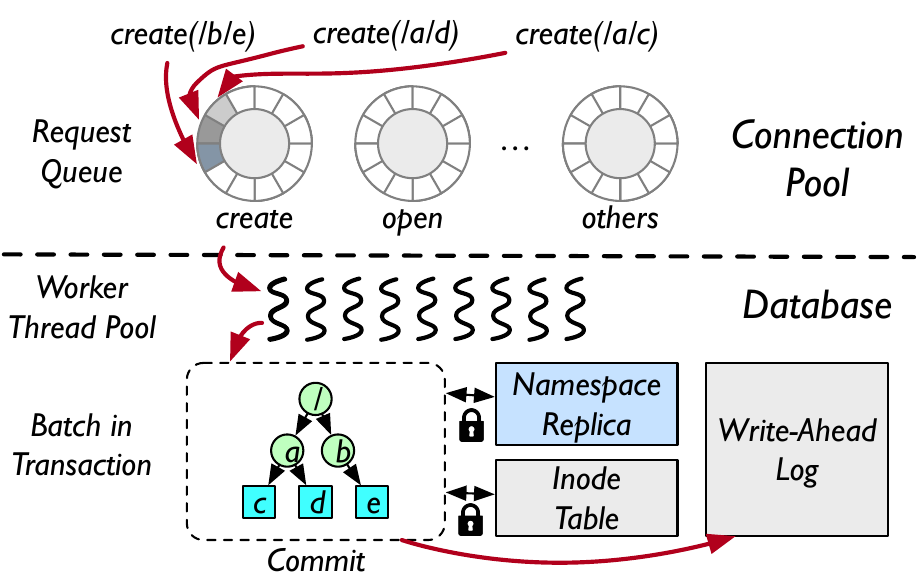}
    \vspace{-10px}
    \caption{\textbf{Concurrent request merging overview.}}
    \label{fig:metadata-server-overview}
    \vspace{-15px}
\end{figure}

\autoref{fig:metadata-server-overview} illustrates the metadata servers' request-handling mechanism, which employs concurrent request merging.
Each MNode initializes a fixed number of database worker threads to serve as the backend for metadata storage and prepares a connection pool.
The connection pool accepts incoming requests and puts them into request queues according to the request type.
An idle worker thread retrieves a queue and executes all requests in the queue in a single database transaction, with the following optimizations.

\paragraph{Lock coalescing.}
During path resolution, the worker thread acquires shared locks for all directories along the path to maintain path validity during operations,
akin to the implementation of VFS and existing DFS.
Prior research demonstrates that lock overhead can be significant even without actual blocking~\cite{CFS}.
{\sys} mitigates this through lock coalescing, combining lock acquisition and release operations at per-batch granularity to reduce overhead.

Due to the tree-structured nature of the file system namespace, requests can share common near-root path prefixes.
As the request queue accumulates multiple operations, the worker coalesces shared path prefixes and eliminates redundant lock acquisitions.
In \autoref{fig:metadata-server-overview}, the three \emph{create} operations each walk two directories and one file.
Rather than acquiring nine locks separately, the worker eliminates redundant lock acquisitions and acquires only six locks instead.

\paragraph{Write-ahead-log coalescing.}
Operations such as \emph{mkdir}, \emph{create}, and \emph{close} modify the inode table.
To maintain file system metadata consistency, DFSs typically warp each operation into a separate transaction to persist in atomic~\cite{Niazi2017HopsFS,CFS,Lv2022InfiniFS}.
When a transaction commits, it synchronously appends the write-ahead log, leading to small writes that are unfriendly for storage.
In {\sys}, as concurrent operations are batched into a single transaction, the worker coalesces small log appends into larger ones, improving the storage efficiency.

\subsection{Reliability and Reconfiguration}

In this section, we discuss how {\sys} supports crash consistency, high availability, and system reconfiguration.

\paragraph{Crash consistency.}
{\sys} adopts write-ahead logging (WAL) to ensure crash consistency and atomicity of operations.
Any persistent updates to the MNodes are first recorded in the WAL before being applied and visible. 
If an MNode fails during path resolution, update, or migration, uncommitted operations will be rolled back, and committed operations will be recovered from the log.
We leverage PostgreSQL's WAL mechanism to support single-node transactions and build a customized two-phase commit protocol upon it to ensure the consistency of operations that span multiple MNodes.
Coordinator failure is treated the same as an MNode failure.

\paragraph{High availability.}
For the high availability of the metadata service, {\sys} supports majority-based replication for MNodes and the coordinator.
Each MNode and the coordinator have multiple replicas, with one primary replica serving requests and multiple secondary replicas synchronizing the primary's state.
The state synchronization is achieved by using PostgreSQL's physical streaming replication mechanism to ship the primary's WAL to the secondaries continuously.
Once a primary replica becomes unavailable, {\sys} elects a secondary replica with the longest WAL as the new primary.
{\sys} is available as long as a majority of each MNode's replicas are available.
Note that the majority-based replication is orthogonal to lazy namespace replication in \autoref{sec:lazy-replication}.

\paragraph{Cluster reconfiguration.}
{\sys} adopts consistent hashing to compute inode location and supports cluster reconfiguration (i.e., MNode joining and leaving) accordingly.
Once the cluster needs to be resized, {\sys} migrates involved inodes to/from the added/removed MNodes.
During migration, {\sys} stops serving requests.
{\sys} does not support live migration since it introduces extra overhead for checking migration status on the critical path of accessing inodes.

\section{Implementation of VFS Compatibility}
\label{sec:lookup-pass-through}

Compatibility with the Linux virtual file system (VFS) is important for \emph{easy deployment}.
However, the VFS embeds path resolution and metadata caching logic within the kernel, which hinders our stateless-client design.
To address this, we shortcut VFS path resolution by leveraging the semantics the VFS already provides, enabling users to benefit from {\sys}'s design without invasive kernel modifications.

\paragraph{Basic idea.} The idea behind \emph{VFS shortcut} is simple --- when the VFS invokes \emph{lookup()} method provided by the client module for intermediate directories in a path, the method returns directory attributes with permission 0777 to pass VFS checks, and when the VFS triggers the operation on the last path component, the client module sends the full path to the metadata servers, which perform the actual path resolution and execute the requested operation.
To implement this approach, two challenges need to be addressed: 
\begin{itemize}
    \item \emph{Distinguishing lookup requests to intermediate directories and the final component.}
    The client module returns fake attributes to the former for shortcutting path resolution and real attributes to the latter for correctness, respectively
    
    \item \emph{Avoiding fake attributes being exposed to users.}
    A previously returned fake attribute may be cached in the kernel and exposed to users during subsequent operations, which violates correctness and should be avoided.
\end{itemize}

\paragraph{Distinguish intermediate and final lookups.}

We observe that the existing semantics provided to the \emph{lookup()} method are sufficient to distinguish lookup intentions.
Since Linux kernel 5.7, the VFS sets the global state flag \emph{LOOKUP\_PARENT} during path walk to indicate that the final component has not yet been reached --- a feature designed initially for the kernel audit subsystem~\cite{path-lookup}, and the flag is passed to the \emph{lookup()} method.
If the flag is set, the client module knows that the lookup is for an intermediate directory and returns fake attributes (e.g., $mode=0777$, along with special $uid$ and $gid$ values) to pass VFS checking.

\paragraph{Avoid exposing fake attributes.}
To avoid fake attributes being exposed to users due to cache reuse, the client module leverages the VFS \emph{d\_revalidate()} method. %
We reserve a pair of \emph{uid} and \emph{gid} to identify fake attributes.
Upon a dcache hit, the VFS invokes \emph{d\_revalidate()} method to validate the cached entry.
The client module then checks whether the hit entry is a fake one via \emph{uid} and \emph{gid}, and whether the entry is being used to resolve a final path component via the \emph{LOOKUP\_PARENT} flag.
If both conditions are met, the module fetches the real attributes from the MNode and updates the dcache entry.

\begin{figure}[t]
    \centering
    \includegraphics[width=\linewidth]{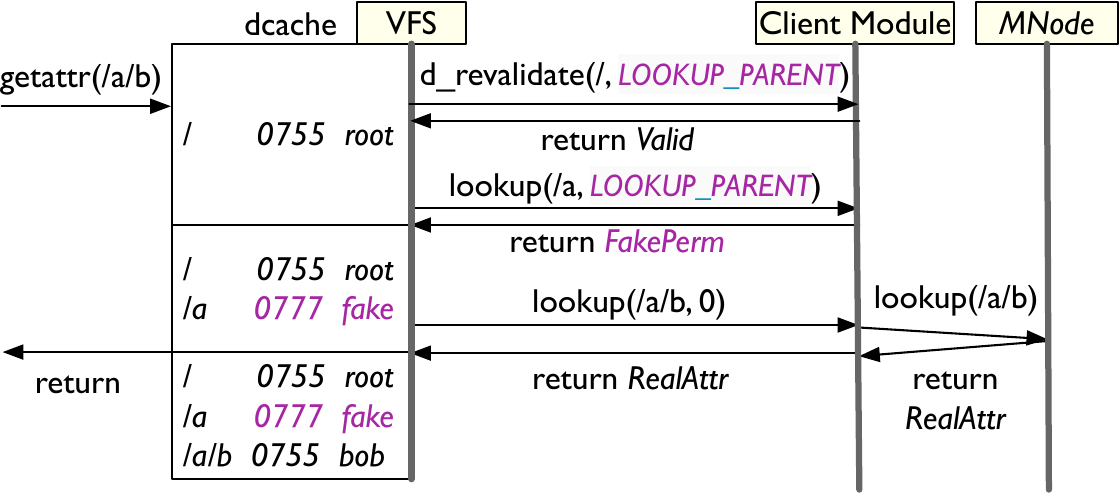}
    \vspace{-15px}
    \caption{\textbf{Workflow of VFS shortcut in {\sys}.} In the figure, The method interfaces and the dcache is simplified for clarity.}
    \label{fig:shadow-lookup}
    \vspace{-15px}
\end{figure}

\paragraph{Example.}
\autoref{fig:shadow-lookup} illustrates an example of VFS shortcut.
During a \emph{getattr} operation for the path \emph{/a/b}, the VFS resolves each component sequentially.
It first looks up \emph{/}, which results in a dcache hit.
The VFS invokes the \emph{d\_revalidate()} method to validate the entry, receiving a positive response.
Then, the VFS looks up \emph{/a}, which misses the dcache.
The \emph{lookup()} method is called with the \emph{LOOKUP\_PARENT} flag set, returning a fake attribute.
Finally, the VFS looks up \emph{/a/b}, which also misses the dcache.
Here, the \emph{lookup()} method is invoked without flags.
The module then issues a remote lookup request with the full path (\emph{/a/b}) to the MNode, which executes the real path checking, executes the lookup operation, and returns the result to the module.
The module returns the result to the VFS, completing the \emph{getattr} operation.

\paragraph{Discussion and limitations.}
Our client-side implementation preserves path resolution correctness and file system operation integrity, as the MNode re-executes all shortcut checks and prevents user exposure to fake attributes.
VFS shortcut has two limitations.
First, symbolic link is not supported because the clients do not follow links.
Second, nested mount points under {\sys} need special handling (i.e., recheck directory permissions when passing the nested mount point), which is not supported yet.

\section{Evaluation}

\begin{figure*}[t]
    \centering
    \begin{minipage}[htp]{1\linewidth}
    \centering
    \includegraphics[width=.7\linewidth]{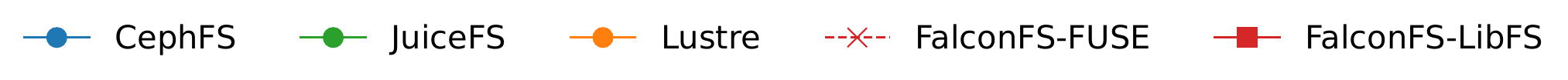}
    \vspace{-5px}
    \end{minipage}
    \begin{minipage}[t]{1\linewidth}
        \begin{minipage}[t]{.195\linewidth}
            \centering
            \includegraphics[width=1\linewidth]{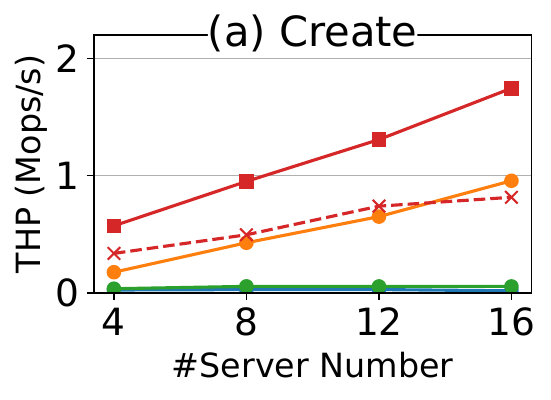}
            \label{fig:throughput_create}
        \end{minipage}
        \begin{minipage}[t]{.195\linewidth}
            \centering
            \includegraphics[width=1\linewidth]{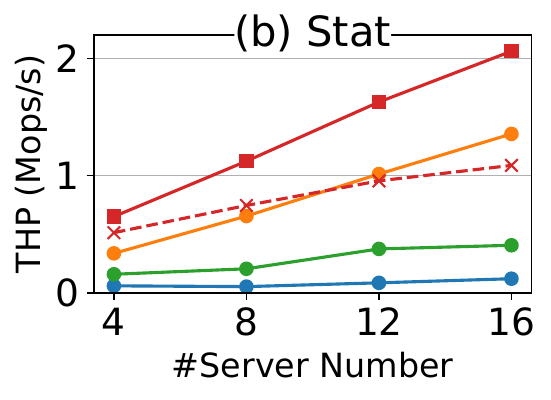}
            \label{fig:throughput_getattr}
        \end{minipage}
        \begin{minipage}[t]{.195\linewidth}
            \centering
            \includegraphics[width=1\linewidth]{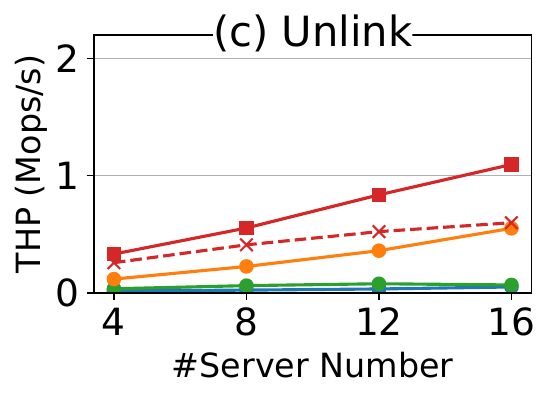}
            \label{fig:throughput_unlink}
        \end{minipage}
        \begin{minipage}[t]{.195\linewidth}
            \centering
            \includegraphics[width=1\linewidth]{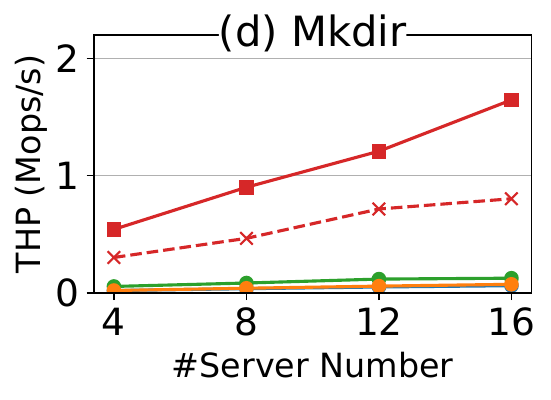}
            \label{fig:throughput_mkdir}
        \end{minipage}
        \begin{minipage}[t]{.195\linewidth}
            \centering
            \includegraphics[width=1\linewidth]{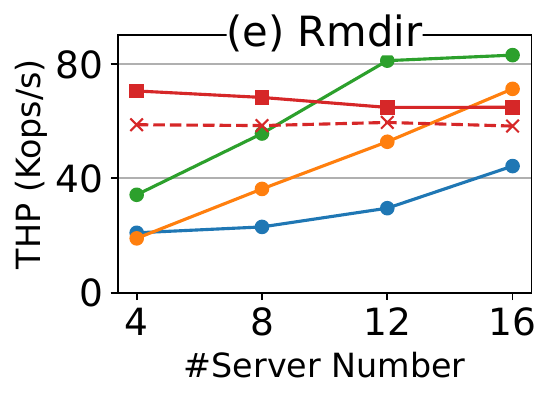}
            \label{fig:throughput_rmdir}
        \end{minipage}
    \end{minipage}
\vspace{-20px}
\caption{\textbf{Performance and scalability of metadata operations.}}
\label{fig:eval-throughput-metadata}
\vspace{-15px}
\end{figure*}

\begin{figure}
    \centering
    \includegraphics[width=1\linewidth]{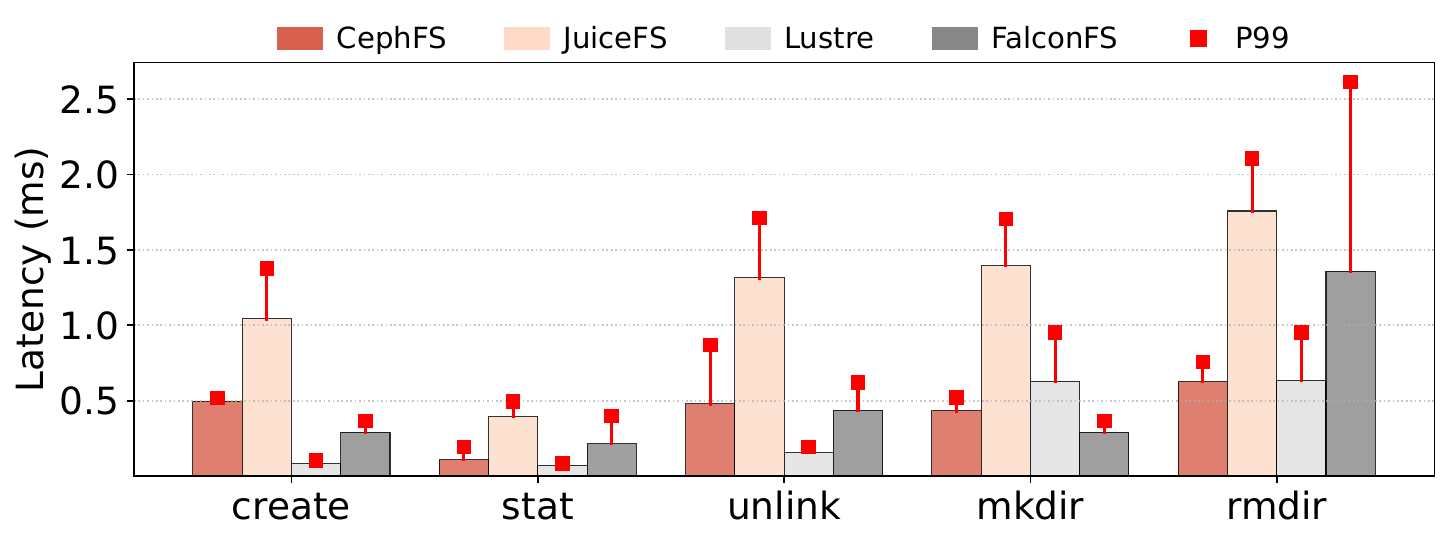}
    \vspace{-20px}
    \caption{\textbf{Average latency of metadata operations.}}
    \label{fig:eval-latency-metadata}
    \vspace{-15px}
\end{figure}

\begin{figure}[ht]
    \centering
    \begin{minipage}{\linewidth}
        \centering
        \includegraphics[width=\linewidth]{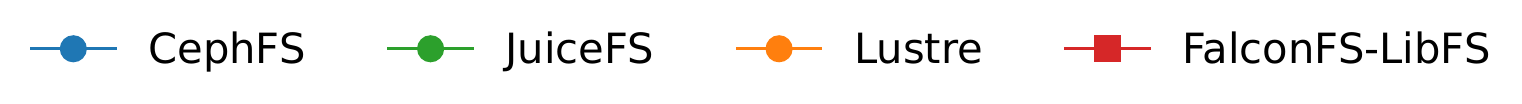}
        \vspace{-18pt}
    \end{minipage}
    \begin{minipage}{.495\linewidth}
        \includegraphics[width=1\linewidth]{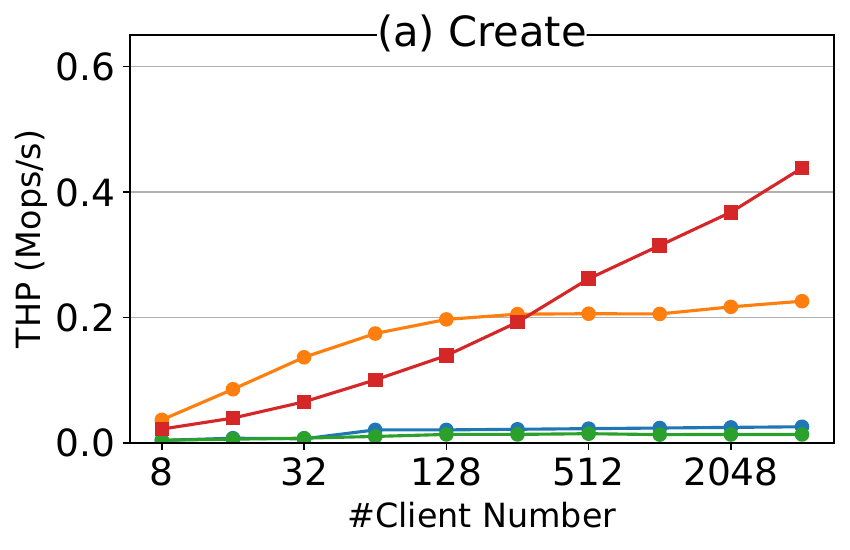}
        \label{fig:eval-client-scalability-create}
    \end{minipage}
    \begin{minipage}{.495\linewidth}
        \includegraphics[width=1\linewidth]{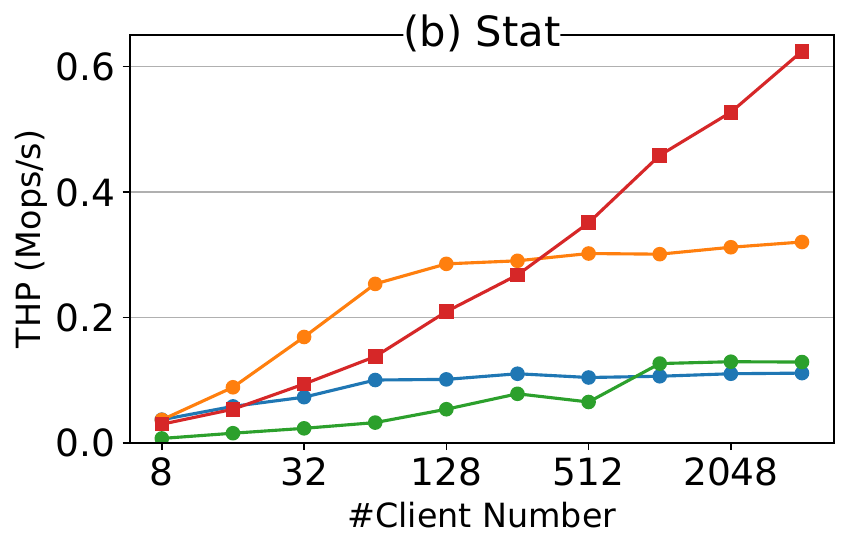}
        \label{fig:eval-client-scalability-stat}
    \end{minipage}
    \vspace{-20px}
    \caption{\textbf{Scalability with regard to concurrent clients.}}
    \label{fig:eval-client-scalability}
    \vspace{-15px}
\end{figure}

We evaluate in this section to present the following results.

\begin{enumerate}
\item {\sys} provides scalable, high-performance metadata operations (\autoref{sec:eval-metadata}) and file IO ( \autoref{sec:eval-file-IO}).

\item {\sys} is robust under adverse conditions like client memory limitations (\autoref{sec:eval-memory}) and load skewness (\autoref{sec:eval-directory-contention}).

\item {\sys} achieves balanced inode distribution across diverse workloads with minimal exception table size (\autoref{sec:eval-load-balance}).

\item The contribution of each design component to overall performance and the impact of unfavorable conditions (\autoref{sec:eval-breakdown}).

\item The performance in real-world DL workloads (\autoref{sec:end-to-end}).
\end{enumerate}

\subsection{Environment Setup}

\textbf{Testbed.} We conduct experiments on a cluster of 13 dual-socket machines (configurations detailed in \autoref{tab:eval-setup}).
To expand the test scale, we abstract each machine into two independent nodes, with each node bound to one socket, one SSD, and one NIC, scaling the testbed to 26 nodes.
We restrict server resources to 4 cores per node to ensure clients can saturate the servers' capabilities.

\begin{table}[t]
\centering
\footnotesize
\caption{Hardware configuration of the cluster.}
\label{tab:eval-setup}
\vspace{-8px}
\begin{tabular}{@{}l|lllll@{}}
\toprule
CPU     & \multicolumn{4}{l}{2 $\times$ Intel Xeon 3.00\,GHz, 12 cores}  \\
Memory  & \multicolumn{4}{l}{16 $\times$ DDR4 2933\,MHz 16\,GB}                      \\
Storage & \multicolumn{4}{l}{2 $\times$ NVMe SSD 960\,GB}                     \\
Network & \multicolumn{4}{l}{2 $\times$ 100\,GbE}           \\ \midrule
\end{tabular}
\vspace{-15px}
\end{table}

\paragraph{Baseline Systems.}
We compare {\sys} with CephFS 12.2.13~\cite{Weil2006Ceph}, JuiceFS 1.2.1~\cite{juicefs} and Lustre 2.15.6~\cite{lustre}.
CephFS is a widely deployed DFS in data centers.
JuiceFS is an open-source DFS targeting AI and data analytics workloads, and we deploy it with TiKV 1.16.1~\cite{TiKV} as its metadata engine and data storage.
Lustre is a high-performance DFS widely used in HPC and data centers.
CephFS is accessed via the libcephfs library due to observed instability in performance when using a VFS mount point.
JuiceFS and Lustre are accessed through VFS mount points, as is {\sys} unless stated otherwise.
{\sys} utilizes a modified FUSE kernel module and library that incorporates the optimizations described in \autoref{sec:lookup-pass-through}.
All DFSs disable metadata and data replication.

\begin{figure*}[ht]
    \centering
    \begin{minipage}{\linewidth}
        \centering
        \includegraphics[width=0.4\linewidth]{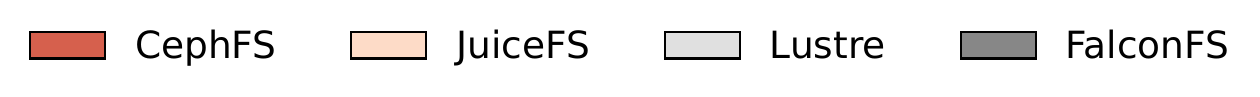}
        \vspace{-5pt}
    \end{minipage}
    \begin{minipage}{.495\linewidth}
        \includegraphics[width=1\linewidth]{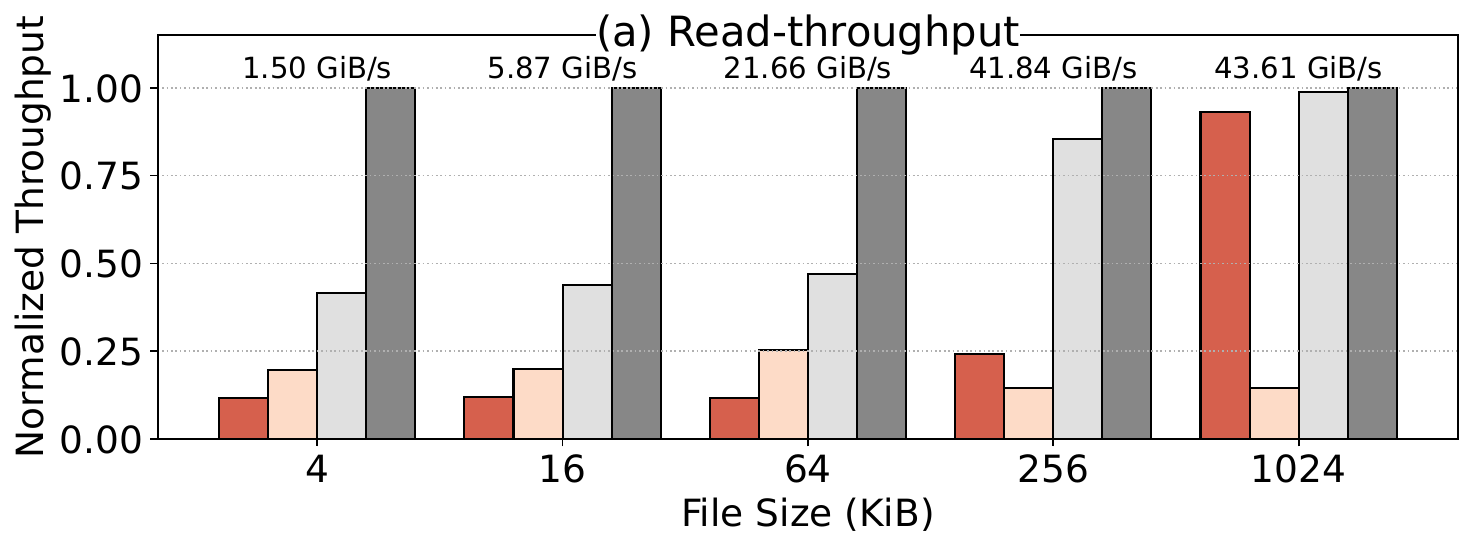}
        \label{fig:eval-throughput-read}
    \end{minipage}
    \begin{minipage}{.495\linewidth}
        \includegraphics[width=1\linewidth]{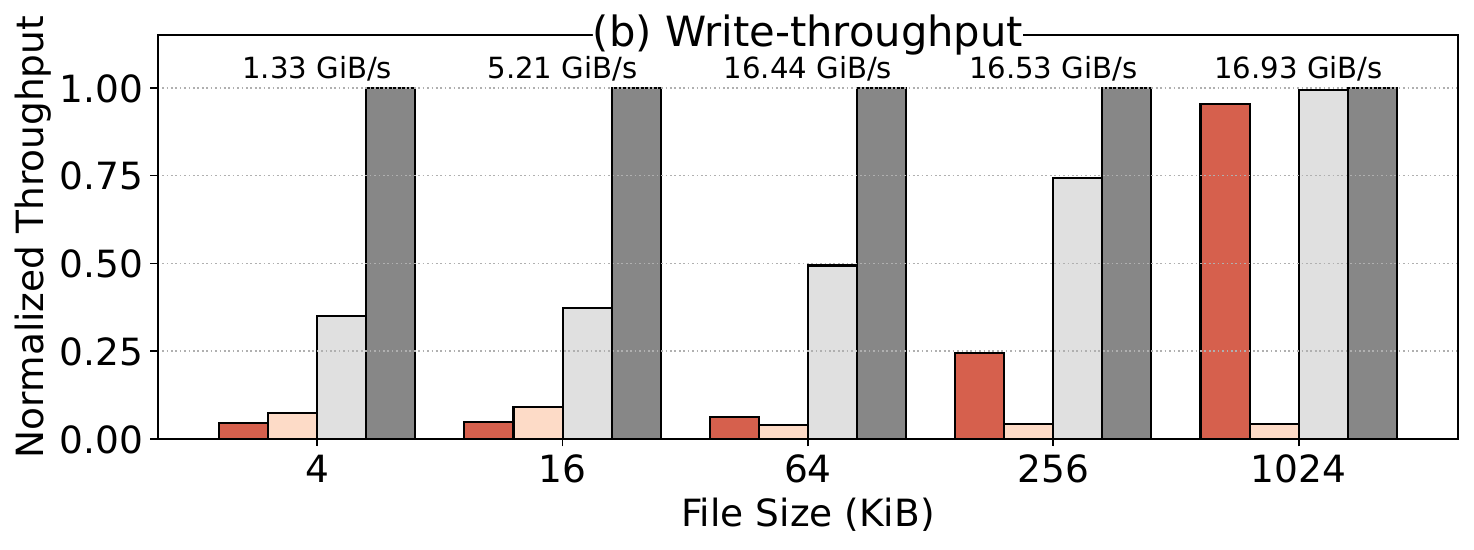}
        \label{fig:eval-throughput-write}
    \end{minipage}
    \vspace{-20px}
    \caption{\textbf{Throughput of file data IO.} Y-axis is the throughput normalized to that of {\sys}. }
    \label{fig:eval-throughput-IO}
    \vspace{-15px}
\end{figure*}

\subsection{Metadata Performance}
\label{sec:eval-metadata}
We first evaluate the performance of individual metadata operations in the best case, where each client accesses its own private directory and all directory lookups hit the client-side cache.
We measure five key metadata operations, namely, \emph{create}, \emph{unlink}, \emph{getattr}, \emph{mkdir} and \emph{rename}.

\paragraph{Throughput scalability.}
We scale the number of metadata servers from 4 to 16 and measure the peak throughput achievable by each file system.
To saturate the metadata servers' capacity, we gradually increase the number of client threads until the throughput no longer increases.
When mounting with FUSE clients, we observe that 13 client nodes are insufficient to saturate {\sys} due to bottlenecks in FUSE, so we present {\sys}'s performance using the LibFS interface, enabling each client node to generate higher concurrency.
We ensure that the FUSE client and the LibFS client generate identical requests to the metadata servers; thus, given sufficient client nodes, FUSE clients would achieve comparable performance.
In production environments, client nodes typically outnumber metadata servers and can fully saturate them.
\autoref{fig:eval-throughput-metadata} shows the results.

For the operations \emph{create} and \emph{unlink}, {\sys} achieves speedups of 0.82--2.26$\times$ on Lustre and higher gains on CephFS and JuiceFS. This performance improvement stems from two factors:
(a) {\sys} does not maintain directories' atime and mtime, eliminating the need to update directory metadata;
(b) Concurrent request merging consolidates and persists multiple write-ahead-logging operations together, enhancing I/O efficiency.
In contrast, JuiceFS and Lustre rely on expensive distributed transactions to update both the file and directory metadata. 
CephFS does not use distributed transactions but logs writes to remote OSDs --- both of which incur significant overhead.

For \emph{getattr}, {\sys} achieves 0.52--0.93$\times$ speedup over Lustre.
The performance gain comes from that (a) concurrent request merging boosts server concurrency and reduces request dispatching overhead, and (b) {\sys}'s stateless-client architecture eliminates the need for acquiring cache coherence locks (e.g., CephFS's capabilities and Lustre's intent locks).

{\sys} demonstrates scalable performance for \emph{mkdir}, due to efficient invalidation-based synchronization.
However, for \emph{rmdir}, {\sys}'s throughput declines as the number of metadata servers increases. 
This is because \emph{rmdir} requires invalidating the directory record and querying child inodes across all servers—an overhead proportional to the cluster size.
In contrast, CephFS, JuiceFS, and Lustre exhibit constant overhead for \emph{rmdir}, so their performance is scalable.

We observe imbalanced CPU utilization across JuiceFS's metadata engine nodes, indicating inefficient load distribution, which explains JuiceFS's poor performance scalability.

\paragraph{Latency.}
\autoref{fig:eval-latency-metadata} presents the latency of metadata operations across different DFSs.
We deployed four metadata servers with a single client thread issuing requests.

While {\sys} demonstrates superior throughput compared to other DFSs, its latency is higher than Lustre's.
This trade-off occurs because {\sys} employs concurrent request merging to batch operations, optimizing throughput at the cost of increased latency.
Besides, {\sys} shows high p99 latency for \emph{rmdir} for broadcasting \emph{invalidation} requests to all MNodes and waiting for the slowest response.
Nevertheless, {\sys}'s latency is comparable to CephFS and better than JuiceFS for operations other than \emph{rmdir}. %

\paragraph{Scalability with regard to concurrent clients.}
\autoref{fig:eval-client-scalability} presents the throughput of metadata operations with an increasing number of client threads, using four metadata servers.
Due to space limitations, we only present the results of \emph{create} and \emph{getattr}.
{\sys}'s throughput scales well for both operations.
When the client number is no more than 256, {\sys}'s throughput is lower than Lustre's due to higher latency.
However, as the number of clients continues to increase, Lustre's performance saturates and {\sys} outperforms it.
{\sys}'s good scalability over client number comes from that
(a) the connection pool allows serving a large number of connections with a few threads, and
(b) the concurrent request merging efficiently batches request executions. 

\subsection{Data Performance}
\label{sec:eval-file-IO}
In this section, we evaluate the performance of small-file access.
We deploy four metadata servers and twelve data nodes, each equipped with one NVMe SSD.
We saturate the DFSs using 2560 client threads distributed across 10 client nodes.
Each thread accesses 1024 pre-created files within its own private directory.
To access a file, a client first opens it with the \emph{O\_DIRECT} flag, reads or writes all data, and then closes the file.
We vary the file size from 4\,KiB to 1\,MiB and report the normalized throughput in \autoref{fig:eval-throughput-IO}.

When the file size is smaller than 256\,KiB, the throughput increases proportionally with the file size, indicating that the metadata operation IOPS is the bottleneck.
When the file size is larger than 256\,KiB, CephFS, Lustre and {\sys}'s throughput hits the SSD bandwidth bottleneck of 43\,GiB/s for read and 16\,GiB/s for write.
Thanks to {\sys}'s higher metadata performance, it outperforms other DFSs in small-file access.
For files no larger than 64\,KiB, {\sys} achieves 7.35--21.23$\times$ speedup over CephFS, 2.94--23.53$\times$ over JuiceFS and 1.12--1.85$\times$ over Lustre.
JuiceFS does not perform well in small file access due to the inefficiency of the data storage.

\subsection{Impact of Client Memory Budget}
\label{sec:eval-memory}

\begin{figure}[t]
    \centering
    \begin{minipage}{\linewidth}
        \includegraphics[width=1\linewidth]{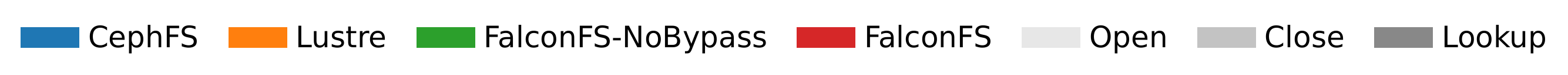}
    \end{minipage}
    \begin{minipage}{.41\linewidth}
        \includegraphics[width=1\linewidth]{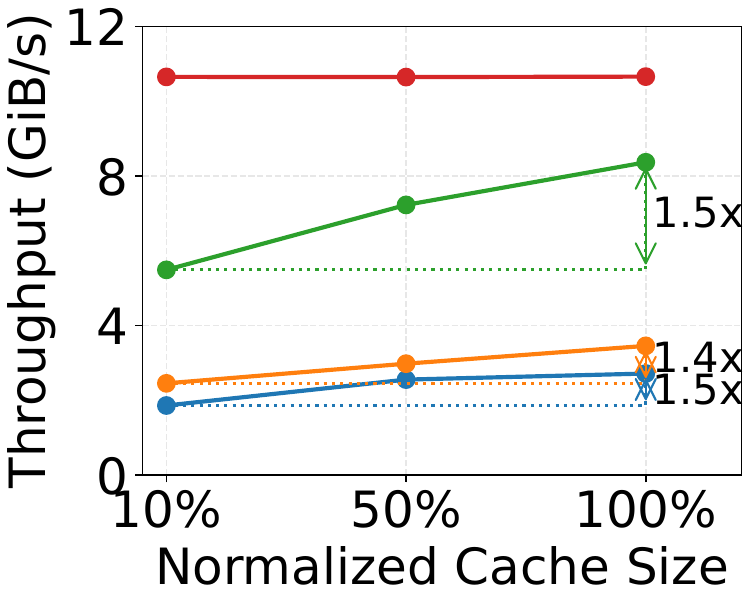}
        \vspace{-15px}
        \subcaption{\textbf{Throughput.}}
        \label{fig:eval-memory-budget-a}
    \end{minipage}
    \begin{minipage}{.58\linewidth}
        \includegraphics[width=1\linewidth]{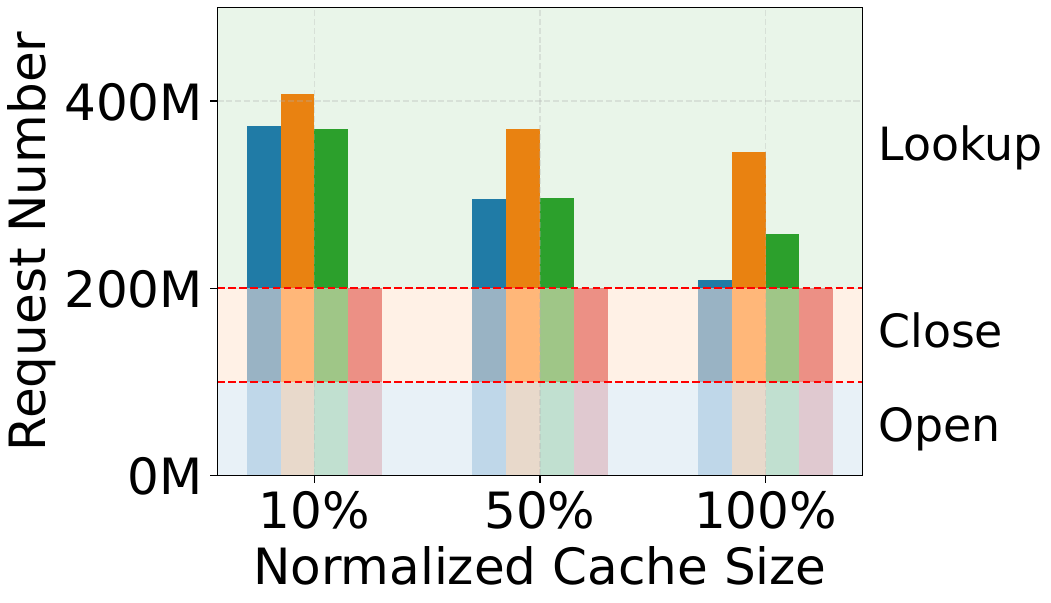}
        \vspace{-15px}
        \subcaption{\textbf{Request number.}}
        \label{fig:eval-memory-budget-b}
    \end{minipage}
    \vspace{-10px}
    \caption{\textbf{Random file traversal in a large directory tree.} The x-axis represents the ratio of client metadata cache size to the total size of all directories' inodes and dentries.}
    \label{fig:eval-memory-budget}
    \vspace{-15px}
\end{figure}

In this section, we evaluate the impact of client memory budget on DFS performance under typical DL training workloads --- specifically, random file traversal in a large directory tree.
We initialize an 8-level directory tree structure where each intermediate directory contains ten subdirectories and each last-level directory contains ten 64\,KiB files.
This configuration yields a total of 11.1 million directories and 100 million files.
Each DFS runs four metadata servers and twelve data nodes.
Ten client nodes, each running a 256-thread client process, read all files in independent random orders.

We limit the metadata cache size on each client node based on the ratio of the total size of all directories' inodes and dentries.
For CephFS, we set the \emph{ceph.conf} parameter \emph{client\_cache\_size} to enforce this limit.
For other DFSs, we use Control Group v2 to restrict the cache size.
The cgroup monitors the process's userspace and kernel memory usage, and reclaims kernel objects (e.g., dentries and inodes) when memory consumption exceeds the threshold.

In addition to CephFS and Lustre, we evaluate {\sys}-NoBypass, a variant of {\sys} without the VFS shortcut, to highlight the benefit of the client-stateless design.
{\sys}-NoBypass relies on the VFS dentry and inode caches to perform client-side path resolution.
We omit the results for JuiceFS, as its throughput drops to zero before completing the initialization of the directory tree.

\autoref{fig:eval-memory-budget-a} shows the throughput of each DFS under different memory budgets, and \autoref{fig:eval-memory-budget-b} presents the composition of metadata requests.
Notably, while Lustre and {\sys} explicitly send \emph{open} requests to open files, CephFS sends \emph{lookup} requests for file \emph{open}.
For simplicity, we count CephFS's \emph{lookup} requests to files as \emph{open} in \autoref{fig:eval-memory-budget-b}.

We make the following observations:
    First, the performance of stateful-client DFSs, including CephFS, Lustre, and {\sys}-NoBypass, is sensitive to the client memory budget.
    There is a 1.4--1.5$\times$ performance gap between the 10\% and 100\% memory budget configurations.
    When the memory budget is constrained, fewer directories can be cached on the client-side, leading to more frequent lookups, which increase the number of requests per file access and degrade throughput.
    
    Second, {\sys} achieves high performance even under tight memory budgets and outperforms {\sys}-NoBypass, demonstrating that stateless-client design effectively boosts performance.
    As shown in \autoref{fig:eval-memory-budget-b}, {\sys} generates a constant number of requests to the metadata servers as the cache size varies.
    Compared with {\sys}-NoBypass, {\sys} reduces the number of metadata requests by 22.7\%--45.9\%, and improves the throughput by 0.24--0.94$\times$.
    Compared with CephFS and Lustre, {\sys} improves the throughput by 2.92--4.72$\times$ and 2.08--3.34$\times$ respectively.

Notably, even with a 100\% cache, {\sys}-NoBypass is still 19.4\% slower than {\sys}.
Although this cache is large enough to hold all directory inodes in the workload, file inodes contend for the cache.
Consequently, directory lookups still frequently miss the cache and generate remote requests.

\subsection{Impact of Transient Skewness}
\label{sec:eval-directory-contention}

\begin{figure}[t]
    \centering
    \begin{minipage}{\linewidth}
        \centering
        \includegraphics[width=0.8\linewidth]{figures/burst_io_legend.pdf}
        \vspace{-5pt}
    \end{minipage}
    \begin{minipage}{.495\linewidth}
        \centering
        \includegraphics[width=1\linewidth]{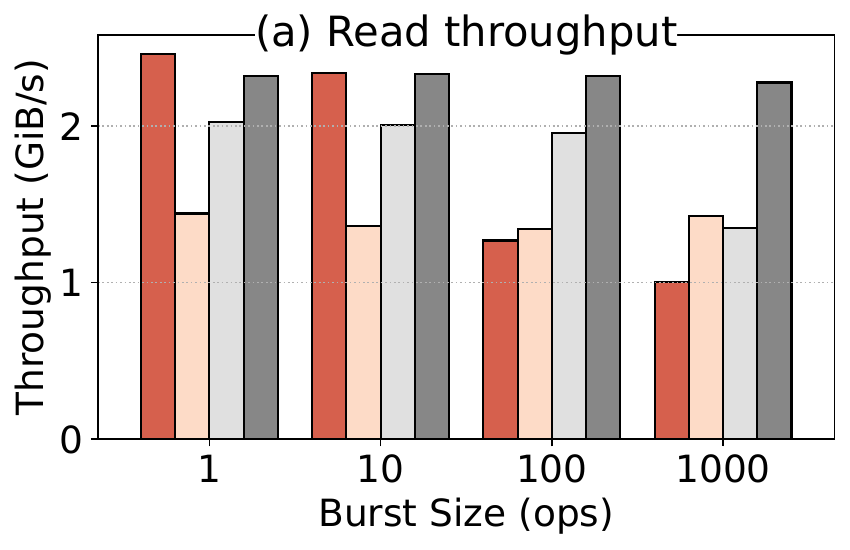}
        \label{fig:eval-burst-read}
    \end{minipage}
    \begin{minipage}{.495\linewidth}
        \centering
        \includegraphics[width=1\linewidth]{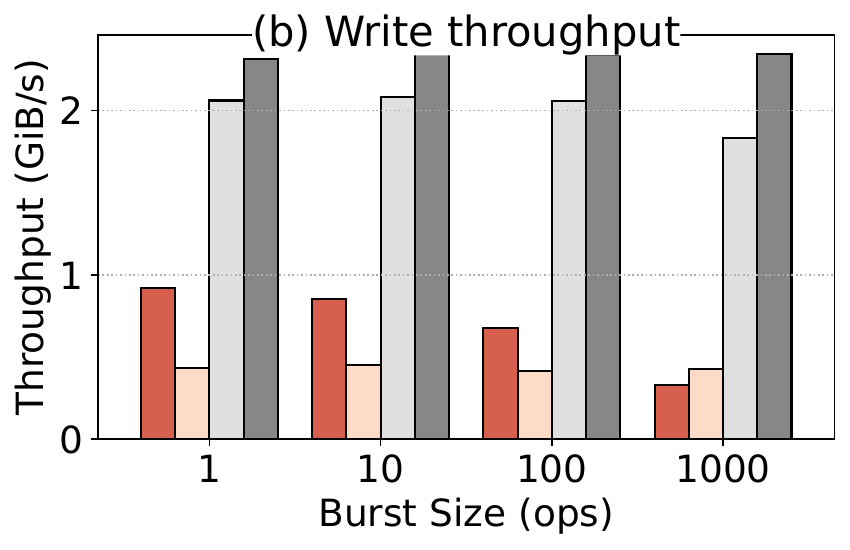}
        \label{fig:eval-burst-write}
    \end{minipage}
    \vspace{-20px}
    \caption{\textbf{Throughput of burst file IO.}}
    \label{fig:eval-burst-IO}
    \vspace{-10px}
\end{figure}

We evaluate the impact of transient skewness on DFS performance, a common access pattern in DL labeling workloads (\autoref{sec:server-concurrency}).
We deploy four metadata servers and twelve data nodes, and use a single 256-thread client node to access pre-created 64\,KiB files in bursts.
A burst is defined as a sequence of accesses to files within the same directory, with adjacent bursts targeting different directories.
\autoref{fig:eval-burst-IO} presents the results.

We observe a degradation in the read and write performance of CephFS and Lustre, as the burst size increases. This occurs because large bursts cause instantaneous load imbalance across the metadata servers.
In contrast, {\sys} does not suffer from large bursts, as it evenly distributes the metadata of files within the same directory, achieving good scalability.
JuiceFS's performance also does not degrade with increasing burst size, as there is a constant load imbalance among its metadata engine nodes.

\subsection{Load Balance in Real Workloads}
\label{sec:eval-load-balance}

\begin{table}[t]
\centering
\footnotesize
\setlength{\tabcolsep}{3pt} 
\caption{\textbf{File and directory inode distribution of various directory structures over 16 metadata servers.}}
\vspace{-8px}
\label{tab:datasets}
    \begin{tabular}{lccccc}
    \toprule
                             & \multirow{2}{*}{inode \#} & \multicolumn{2}{c}{inode distribution} & \multicolumn{2}{c}{exception entry \#}    \\
                             &                   & max       & min      & path-walk & overriding \\
    \midrule
    Labeling task            &   33320           &  6.99\%   &  5.30\%   &    0       &      0       \\
    \hline
    ImageNet~\cite{imagenet-object-localization-challenge}         &   2027728        &  6.29\%   &  6.21\%   &    0       &      0       \\
    KITTI~\cite{Geiger2012CVPR}            &   15003          &  7.01\%   &  5.47\%   &    0       &      0       \\
    Cityscapes~\cite{Cordts2016Cityscapes}       &   20022          &  6.30\%   &  6.22\%   &    0       &      0       \\
    CelebA~\cite{liu2015faceattributes}           &   202599         &  6.54\%   &  6.95\%   &    0       &      0       \\
    SVHN~\cite{Netzer2011ReadingDI}             &   33404          &  6.77\%   &  5.76\%   &    0       &      0       \\
    CUB-200-2011~\cite{WahCUB_200_2011}     &   12003          &  6.68\%   &  5.95\%   &    0       &      0       \\
    \hline
    Linux-6.8 code           &   88936          &  6.49\%   &  5.96\%   &    2       &      0       \\
    FSL homes~\cite{snia-trace-static-5228}                &   655177           & 6.83\%   &  5.45\%   &    1       &      0       \\
    \bottomrule
    \end{tabular}
\setlength{\tabcolsep}{6pt} 
\vspace{-12px}
\end{table}

In this section, we demonstrate that hybrid metadata indexing achieves a balanced distribution of inodes across diverse directory structures, with only a small portion of filenames requiring special treatment.
We evaluate inode distribution on both DL workloads and general-purpose workloads.
For DL workloads, we analyze a dataset collected from {\company}'s production environment, as well as six popular open-source image datasets used in deep learning.
We select these open-source datasets by listing the most popular image datasets on a dataset summary website~\cite{paperswithcode}, and choose the first six datasets containing more than 10,000 files.
For general-purpose workloads, we select the Linux 6.8 source tree and FSL home traces~\cite{snia-trace-static-5228}, the latter being a snapshot of students' home directories from a shared NFS on a university campus.

\autoref{tab:datasets} summarizes, for each workload, the number of files, the maximum and minimum ratios of inodes on metadata servers, and the number of exception entries utilized to achieve the distribution.
As shown in the table, most workloads exhibit a small \emph{max}--\emph{min} gap with zero exception entries, indicating that filename hashing alone is sufficient to achieve balanced inode distribution in these cases.
This is because such workloads --- typically datasets --- have a large directory size, which is friendly to filename hashing.

We further examine the three workloads that require exception entries.
The ``Linux 6.8 code tree'' contains many files with identical names.
However, applying path-walk redirection to the two most frequent filenames (i.e., ``Makefile'' and ``Kconfig'') suffices to balance the distribution.
These two filenames occur 2,945 and 1,690 times, respectively, accounting for 5.55\% of all files in total.
In the FSL home traces, {\sys} achieves balanced load after applying path-walk redirection to the most frequent filename, which appears 8,112 times and accounts for 1.24\% of all files.

\subsection{Performance Analysis}
\label{sec:eval-breakdown}

\paragraph{Design contributions.}

\begin{figure}[t]
    \centering
    \begin{minipage}{0.49\linewidth}
        \centering
        \includegraphics[width=1\linewidth]{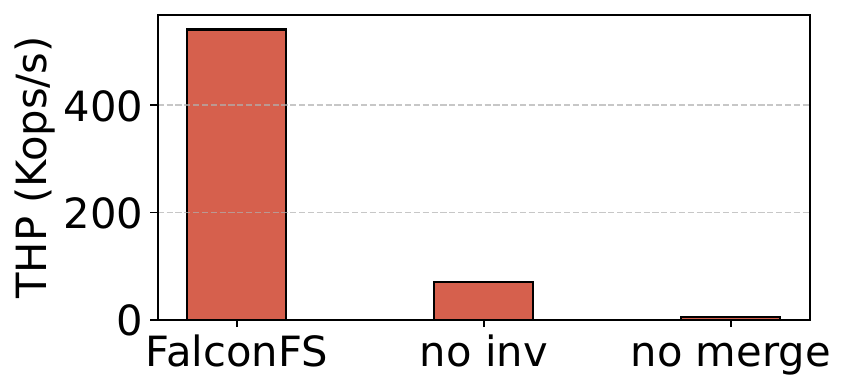}
        \vspace{-15px}
        \subcaption{\textbf{Contribution breakdown.}}
        \label{fig:eval-breakdown-contribution}
    \end{minipage}
    \begin{minipage}{0.49\linewidth}
        \centering
        \includegraphics[width=1\linewidth]{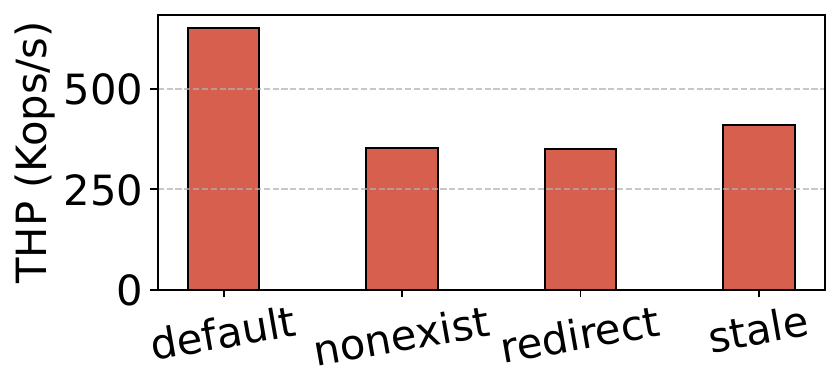}
        \vspace{-15px}
        \subcaption{\textbf{Corner-case analysis.}}
        \label{fig:eval-breakdown-corner-case}
    \end{minipage}
    \vspace{-5px}
    \caption{\textbf{Performance analysis.}}
    \label{fig:eval-breakdown}
    \vspace{-15px}
\end{figure}

The contribution of the stateless-client architecture to overall performance is demonstrated by comparing {\sys} with {\sys}-NoBypass in \autoref{sec:eval-memory}.
In this experiment, we further analyze other design configurations by evaluating three setups: the full \emph{{\sys}}, \emph{no inv}, and \emph{no merge}, incrementally reducing the design features.
The \emph{no inv} configuration disables invalidation-based synchronization, wrapping \emph{mkdir} operations in a distributed transaction to atomically create all dentry replicas across all MNodes.
The \emph{no merge} setup disables concurrent request merging in addition to the changes in \emph{no inv}, requiring worker threads to fetch and execute requests one at a time.
For this evaluation, we deploy four MNodes and use LibFS clients to saturate them, with each client accessing its own private directory.
\autoref{fig:eval-breakdown-contribution} presents the peak throughput of the \emph{mkdir} operation.

Compared to the full \emph{{\sys}}, \emph{no inv} decreases throughput by 86.9\%, as the distributed transaction requires multiple rounds of broadcasts, incurring significant overhead.
\emph{no merge} reduces throughput by an additional 91.8\% due to increased per-request overhead.

\paragraph{Corner case analysis.}
In most cases, hybrid metadata indexing enables one-hop operations.
However, there are corner cases that require two hops: (a) operations on non-existent paths, (b) operations on path-walk redirected filenames, and (c) operations issued with a stale exception table.
\autoref{fig:eval-breakdown-corner-case} illustrates how these scenarios affect the performance of the \emph{getattr} operation.
Compared to the one-hop common case, these corner cases result in a 36.8\%–49.6\% decrease in performance due to the additional hop.

\subsection{End-to-End Performance}
\label{sec:end-to-end}

We evaluate the end-to-end performance of DFSs in DL workloads for both labeling and training tasks.
Each DFS has four metadata servers and twelve data nodes.

\begin{figure}[t]
    \centering
    \begin{minipage}{0.49\linewidth}
        \includegraphics[width=1\linewidth]{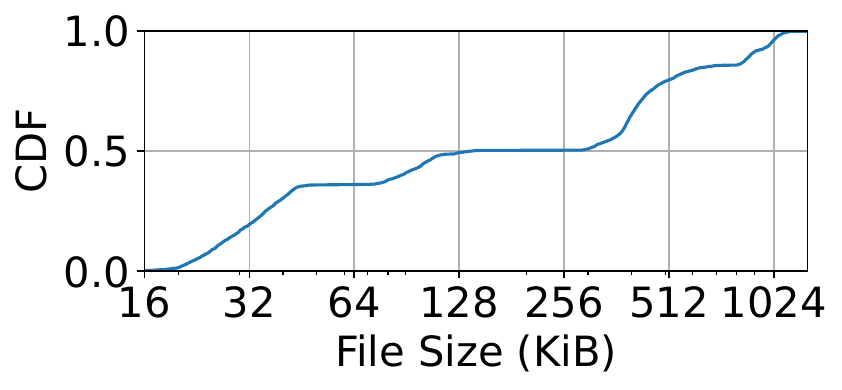}
        \vspace{-15px}
        \subcaption{\textbf{Distribution of file size.}}
        \label{fig:eval-file-size-cdf}
    \end{minipage}
    \begin{minipage}{.49\linewidth}
        \includegraphics[width=1\linewidth]{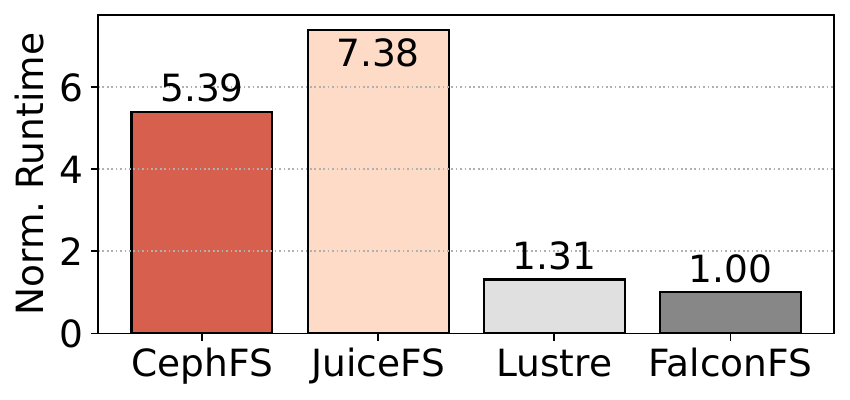}
        \vspace{-15px}
        \subcaption{\textbf{Normalized trace runtime.}}
        \label{fig:eval-labeling-runtime}
    \end{minipage}
    \vspace{-8px}
    \caption{\textbf{File size pattern and runtime for labeling task replay.}}
    \label{fig:eval-labeling-trace}
    \vspace{-12px}
\end{figure}

\paragraph{The labeling task.}
We replay a trace from {\company}'s labeling cluster.
In this trace, labeling tasks read raw images from the DFS and write segmented images back to the DFS.
\autoref{fig:eval-file-size-cdf} shows the distribution of file sizes in the trace, and \autoref{fig:eval-labeling-runtime} presents the runtime of the trace replay.
Although we do not replay the computation, the replay runtime closely approximates the end-to-end runtime, as computation overlaps with I/O, and I/O is the bottleneck.
Compared to other DFSs, {\sys} reduces the runtime by 23.8\%–86.4\%.

\begin{figure}[t]
    \centering
    \begin{minipage}{0.49\linewidth}
        \includegraphics[width=1\linewidth]{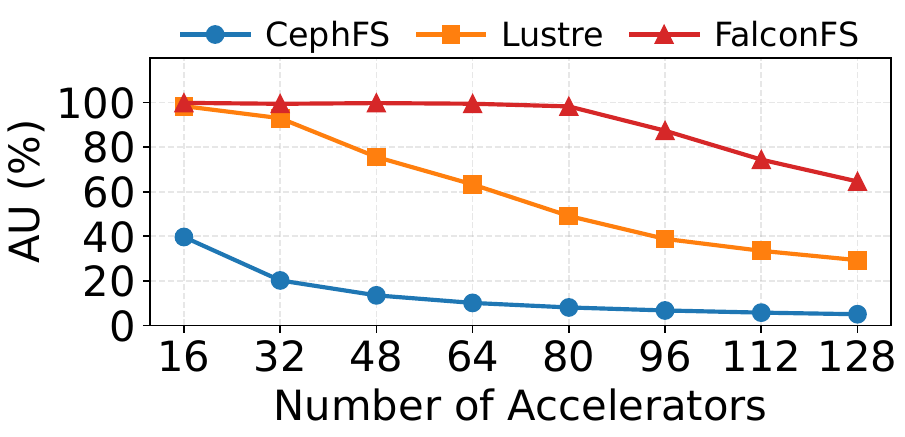}
        \vspace{-15px}
        \subcaption{\textbf{Accelerator utilization.}}
        \label{fig:eval-mlperf-utilization}
    \end{minipage}
    \begin{minipage}{0.49\linewidth}
        \includegraphics[width=1\linewidth]{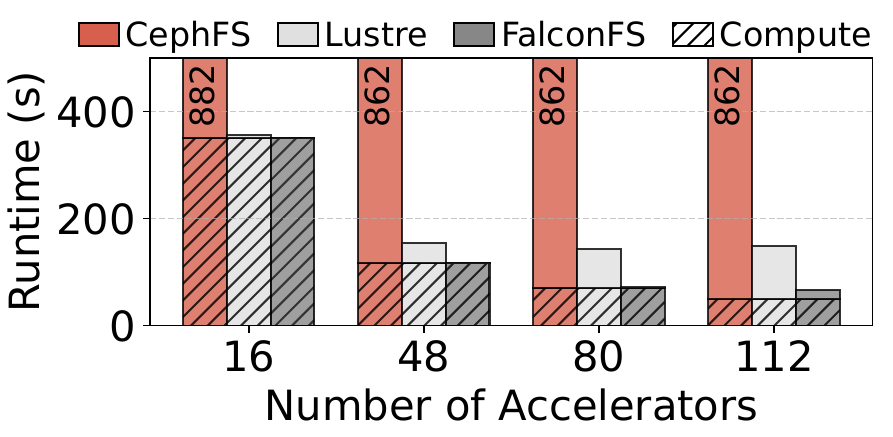}
        \vspace{-15px}
        \subcaption{\textbf{Runtime breakdown.}}
        \label{fig:eval-mlperf-runtime}
    \end{minipage}
    \vspace{-8px}
    \caption{\textbf{Accelerator utilization and runtime breakdown for Resnet-50 model training.} In \autoref{fig:eval-mlperf-runtime}, the stripes mark the computation time, and the bars above the stripes represent the time waiting for I/O.}
    \label{fig:eval-mlperf}
    \vspace{-15px}
\end{figure}

\paragraph{The training task.}
We evaluate the training performance with the MLPerf Storage Benchmark~\cite{mlperf}.
The benchmark is configured to simulate training a ResNet-50 model on 10 million files distributed across 1 million directories, with each file sized at 112\,KiB.
The total dataset size is 1,064 \,GiB, and files are read using direct I/O.
\autoref{fig:eval-mlperf-utilization} shows the accelerator utilization (AU) of each DFS as the number of GPUs increases.
JuiceFS is omitted because its throughput drops to zero during dataset initialization.
Taking 90\% AU as the threshold, {\sys} supports up to 80 GPUs, while Lustre supports only 32 GPUs, and CephFS does not meet the threshold.
With 80 to 128 GPUs, {\sys} achieves training throughput speedups of 11.09--11.81$\times$ over CephFS and 0.99--1.23$\times$ over Lustre.
\autoref{fig:eval-mlperf-runtime} presents the runtime breakdown of the training task.
Due to {\sys}'s high performance for random small-file access, its I/O overlaps with computation, spending significantly less time waiting for I/O compared to other DFSs, thereby reducing the overall training runtime.

\section{Related Works}%
\label{sec:relatedworks}

\paragraph{Path resolution optimizations.}
Path resolution overhead has drawn research attention for a long time.
In the context of local file systems, a series of studies propose optimizations like full-path indexing~\cite{10.1145/2485732.2485741,10.1145/2815400.2815405,FlatFS} and VFS modifications~\cite{216892}.
These approaches optimize local data structures and cannot be directly applied to distributed file systems (DFSs).

To accelerate path resolution, DFSs typically adopt client-side metadata caching~\cite{nfsv4,Ren2014IndexFS,Li2017LocoFS,Weil2006Ceph,lustre,BeeGFS,Lv2022InfiniFS}. InfiniFS~\cite{Lv2022InfiniFS} reduces the cache misses penalty by resolving multiple path components in parallel; however, it cannot mitigate request amplification.
Giraffa~\cite{Giraffa} and CalvinFS~\cite{CalvinFS} locate inodes by full path hashing, which makes directory renaming hard to implement.
HDFS~\cite{Shvachko2010The-Hadoop} performs path resolution on a centralized namenode and thus has scalability issues.
HopsFS~\cite{Niazi2017HopsFS} performs path resolution at a proxy layer, which looks up all path components in parallel from a distributed database, but still suffers from constant request amplification.
Our approach differs in that {\sys} addresses scalability, request amplification, and metadata indexing issues through a client-stateless architecture and hybrid metadata indexing.
Mantle~\cite{Mantle}, a concurrent work of {\sys}, adopts a similar client-stateless metadata service to object storage services.
While {\sys} scales out path resolution to all NModes, Mantle resolves path in a per-namespace single node and has scalability issues.

\paragraph{Storage systems for deep learning.}
Previous studies propose various data loading frameworks for DL training tasks~\cite{OneAccess,Revamper,CoorDL,Quiver,SiloD}.
These works optimize DL data loading by unifying data access, reusing pre-processed data, and leveraging data caching, etc.
These works are task-specific and can be deployed on top of {\sys}.
DIESEL~\cite{DIESEL} is a DFS designed for DL training tasks.
Its design is based on the assumption that the dataset's directory tree is small enough to be cached on every client, whereas {\sys} makes the opposite assumption and focuses on eliminating client-side caching.
3FS~\cite{Fire-Flyer} is a recent DFS for AI workloads.
Unlike {\sys}, it is optimized for large data access and focuses on optimizing the data path, while {\sys} optimizes the metadata architecture for small-file performance.

\section{Conclusion}
We propose {\sys}, a distributed file system with a client-stateless architecture for DL workloads.
Evaluations show that {\sys} achieves up to 4.72$\times$ better throughput of small file random access and up to 11.81$\times$ higher GPU utilization in deep learning model training over CephFS and Lustre. 
\section*{Acknowledgements}
We sincerely thank our shepherd, Ken Birman, and the anonymous reviewers for their constructive comments and insightful suggestions.
This work is supported in part by the National Natural Science Foundation of China (No. 62132014), the Fundamental Research Funds for the Central Universities, the Fundamental and Interdisciplinary Disciplines Breakthrough Plan of the Ministry of Education of China (JYB2025XDXM113), and Huawei Technologies.
Mingkai Dong (\url{mingkaidong@sjtu.edu.cn}) and Junbin Kang (\url{kangjunbin1@huawei.com}) are the corresponding authors.

\bibliographystyle{plain}
\bibliography{references}

\appendix
\section{Appendix}

\subsection{Theoretical Analysis}
\label{sec:theoretical-analysis}

We demonstrate that hybrid metadata indexing (\autoref{sec:filename-direct-indexing}) achieves an even distribution of inodes with at most $O(nlogn)$ exception table entries --- not only for DL workloads but also for arbitrary directory structures, where $n$ denotes the number of MNodes.
We start our discussion with strong assumptions on the directory structure and progressively relax them.

\paragraph{Many filenames, identical frequency.}
First, we assume that the file system namespace contains significantly more unique filenames than MNodes, with each filename appearing an equal number of times.
Under this condition, filename hashing ensures a statistically even distribution of inodes across MNodes, as dictated by the law of large numbers.

\paragraph{Many filenames, varying frequency.}
Then we remove the assumption that all filenames appear with equal frequency.
We demonstrate that by applying \emph{path-walk redirection} to the $O(nlogn)$ most frequent filenames and applying filename hashing to the remainder, an even distribution of inodes across MNodes can be achieved --- regardless of the underlying filename frequency distribution.

Our proof builds upon a theorem from caching literature~\cite{10.1145/2038916.2038939,Jin2017NetCache}, which states:
for $m$ objects randomly partitioned across $n$ nodes with a total query load of $n\cdot t$, if a cache absorbs all queries to the hottest $O (nlogn)$ items, then no node exceeds $t$ load with high probability, independent of the query distribution. We adapt this through constructive proof.

Consider $n\cdot t$ files with $m$ distinct filenames, randomly partitioned across $n$ nodes via filename hashing, and a query load accessing each file uniformly.
Now we think of the filenames as the objects in the theorem.
The query load on each filename is proportional to the number of files with that filename.
The theorem guarantees that after removing queries for the hottest $O (nlogn)$ filenames, the remaining load is evenly distributed across nodes.
It indicates that files not among these hottest $O(nlogn)$ filenames must themselves be evenly distributed across nodes.

Now that the theorem guarantees that files not among the hottest $O(nlogn)$ filenames are evenly distributed across nodes and that we apply \emph{path-walk redirection} to the $O(nlogn)$ most frequent filenames to ensure their even distribution, the entire namespace must be evenly distributed.

\paragraph{A few filenames, varying frequency.}
Finally, we relax the assumption that filenames significantly outnumber MNodes, considering instead the case where only $O(n)$ distinct filenames exist in the namespace.
A trivial solution for achieving even inode distribution with at most $O(n)$ exception entries would be to apply \emph{path-walk redirection} to all filenames, thus completing our theoretical proof.

In practice, we avoid \emph{path-walk redirection} since it introduces an additional hop for file operations.
Instead, our load balancing algorithm (\autoref{sec:scheduling}) prioritizes \emph{overriding redirection} over \emph{path-walk redirection}, resorting to the latter only when necessary.

\subsection{Orthogonal Task-Level Optimizations}
Previous studies have proposed task-level optimizations that change the way in which data is shuffled to make the I/O pattern more friendly to the DFS~\cite{DIESEL,Quiver}.
Specifically, they group data objects into partitions, and shuffle the order of partitions and the order of objects in each partition separately for each epoch, in order to reduce the scope of the random access footprint.

{\sys}'s optimization is orthogonal to these task-level optimizations.
While the task-level optimizations require engineering efforts on the training framework layer to implement and constrain the way in which data are shuffled,
{\sys} satisfies the data demand of training tasks through filesystem-level optimization, which is transparent to upper-layer tasks and leaves sufficient room for the tasks to conduct orthogonal designs and optimizations.

\end{document}